\documentclass[longauth]{aa} 

\usepackage{graphicx}
\usepackage{txfonts}
\usepackage{soul}
\usepackage{natbib}
\usepackage{pdfpages}
\usepackage[colorlinks,urlcolor=cyan,citecolor=blue,linkcolor=blue]{hyperref}
\usepackage{amssymb,amsmath}
\usepackage{array}
\usepackage{booktabs}
\providecommand{\bjdtdb}{\ensuremath{\rm {BJD_{TDB}}}}

\providecommand{\fave}{\langle F \rangle}
\providecommand{\msun}{\ensuremath{\rm\,M_\odot}}
\providecommand{\rsun}{\ensuremath{\rm\,R_\odot}}
\providecommand{\lsun}{\ensuremath{\rm\,L_\odot}}

\providecommand{\me}{\ensuremath{\rm\,M_\oplus}}
\providecommand{\re}{\ensuremath{\rm\,R_\oplus}}
\providecommand{\fluxcgs}{10$^9$ erg s$^{-1}$ cm$^{-2}$}
\providecommand{\tar}{TOI-1680}
\providecommand{\gaia}{\textit{Gaia}}
\providecommand{\TESS}{\textit{TESS}}
\providecommand{\JWST}{\textit{JWST}}

\begin{document}

\title{\TESS\ discovery of a super-Earth orbiting the M-dwarf star TOI-1680}

\author{ M.~Ghachoui\inst{\ref{liege},\ref{oukaimeden}}
\and A.~Soubkiou  \inst{\ref{oukaimeden},\ref{porto-fis},\ref{porto-iace}}
\and R.D.~Wells \inst{\ref{bern}}
\and B.V.~Rackham \inst{\ref{mit_eaps}, \ref{mit_kavli}, \thanks{51 Pegasi b Fellow}}
\and A.~H.M.J.~Triaud \inst{\ref{ubirm}}
\and D.~Sebastian \inst{\ref{ubirm}}
\and S.~Giacalone \inst{\ref{ucb}}
\and K.G.~Stassun \inst{\ref{vandy}}
\and D.R.~Ciardi \inst{\ref{ipac}}
\and K.A.~Collins \inst{\ref{Harvard&Smithsonian}}
\and A.~Liu \inst{\ref{mit_eaps}}
\and Y.~G\'omez~Maqueo~Chew \inst{\ref{ciudad}}
\and M.~Gillon \inst{\ref{liege}}
\and Z.~Benkhaldoun \inst{\ref{oukaimeden}}
\and L.~Delrez \inst{\ref{liege},\ref{liege_star}}
\and J.D.~Eastman \inst{\ref{Harvard&Smithsonian}}
\and O.~Demangeon  \inst{\ref{porto-fis},\ref{porto-iace}}
\and K.~Barkaoui \inst{\ref{liege},\ref{mit_eaps},\ref{iac}}
\and A.~Burdanov \inst{\ref{mit_eaps}}
\and B.-O.~Demory \inst{\ref{bern}}
\and J.~de~Wit \inst{\ref{mit_eaps}}
\and G.~Dransfield \inst{\ref{ubirm}}
\and E.~Ducrot \inst{\ref{paris},\ref{paris_region_fellow}}
\and L.~Garcia \inst{\ref{liege}}
\and M.A.~G\'omez-Mu\~noz \inst{\ref{ciudad1}}
\and M.J.~Hooton \inst{\ref{cavendish}}
\and E.~Jehin \inst{\ref{liege_star}}
\and C.A.~Murray \inst{\ref{cavendish}}
\and P.P. Pedersen \inst{\ref{cavendish}}
\and F.J.~Pozuelos \inst{\ref{iaa}} 
\and D.~Queloz  \inst{\ref{cavendish}}
\and L.~Sabin \inst{\ref{ciudad1}}
\and N.~Schanche \inst{\ref{bern}}
\and M.~Timmermans \inst{\ref{liege}}
\and E.J.~Gonzales \inst{\ref{ucsc}}
\and C.D.~Dressing \inst{\ref{ucb}}
\and C.~Aganze \inst{\ref{ucsd}}
\and A.J.~Burgasser \inst{\ref{ucsd}}
\and R.~Gerasimov \inst{\ref{ucsd}}
\and C.~Hsu \inst{\ref{ucsd},\ref{northwestern}}
\and C.A.~Theissen \inst{\ref{ucsd}}
\and D.~Charbonneau \inst{\ref{Harvard&Smithsonian}}
\and J.M.~Jenkins \inst{\ref{nasa_ames}}
\and D.W.~Latham \inst{\ref{Harvard&Smithsonian}}
\and G. Ricker \inst{\ref{mit_kavli},\ref{DP_MIT}}
\and S.~Seager \inst{\ref{mit_eaps},\ref{DAA_MIT},\ref{DP&KIA&SR}}
\and A.~Shporer \inst{\ref{DP&KIA&SR}}
\and J.D.~Twicken \inst{\ref{SETI}}
\and R.~Vanderspek \inst{\ref{DP&KIA&SR}}
\and J.N.~Winn \inst{\ref{DAS_PU}}
\and K.I.~Collins \inst{\ref{gmu}}
\and A.~Fukui \inst{\ref{Komaba_Tokyo},\ref{iac_38205}}
\and T.~Gan \inst{\ref{Tsinghua_china}}
\and N.~Narita \inst{\ref{Komaba_Tokyo},\ref{Osawa_Tokyo},\ref{iac_38205}}
\and R.P.~Schwarz \inst{\ref{Harvard&Smithsonian}}
}

\institute{Astrobiology Research Unit, Universit\'e de Li\`ege, All\'ee du 6 Ao\^ut 19C, B-4000 Li\`ege, Belgium \label{liege}
\and Oukaimeden Observatory, High Energy Physics and Astrophysics Laboratory, Cadi Ayyad University, Marrakech, Morocco \label{oukaimeden}
\and Departamento de Fisica e Astronomia, Faculdade de Ciencias, Universidade do Porto, Rua do Campo Alegre, 4169-007 Porto, Portugal \label{porto-fis}
\and Instituto de Astrofisica e Ciencias do Espaco, Universidade do Porto, CAUP, Rua das Estrelas, 150-762 Porto, Portugal \label{porto-iace}
\and Center for Space and Habitability, University of Bern, Gesellschaftsstrasse 6, CH-3012, Bern, Switzerland \label{bern}
\and Department of Earth, Atmospheric and Planetary Science, Massachusetts Institute of Technology, 77 Massachusetts Avenue, Cambridge, MA 02139, USA \label{mit_eaps}
\and Kavli Institute for Astrophysics and Space Research, Massachusetts Institute of Technology, Cambridge, MA 02139, USA \label{mit_kavli}
\and 51 Pegasi b Fellow \label{51pegb}
\and School of Physics \& Astronomy, University of Birmingham, Edgbaston, Birmimgham B15 2TT, UK \label{ubirm}
\and Department of Astronomy, University of California Berkeley, Berkeley, CA 94720, USA \label{ucb}
\and Department of Physics and Astronomy, Vanderbilt University, Nashville, TN 37235, USA \label{vandy}
\and NASA Exoplanet Science Institute-Caltech/IPAC, 1200 E. California Blvd, Pasadena, CA 91125 USA \label{ipac}
\and Center for Astrophysics \textbar \ Harvard \& Smithsonian, 60 Garden Street, Cambridge, MA 02138, USA \label{Harvard&Smithsonian}
\and Universidad Nacional Aut\'onoma de M\'exico, Instituto de Astronom\'ia, AP 70-264, Ciudad de M\'exico,  04510, M\'exico \label{ciudad} 
\and Space Sciences, Technologies and Astrophysics Research (STAR) Institute, Universit\'e de Li\`ege, All\'e du 6 Ao\^ut 19C, B-4000 Li\`ege, Belgium \label{liege_star}
\and Instituto de Astrof\'isica de Canarias (IAC), Calle V\'ia L\'actea s/n, 38200, La Laguna, Tenerife, Spain \label{iac}
\and AIM, CEA, CNRS, Universit\'e Paris-Saclay, Universit\'e de Paris, F91191 Gif-sur-Yvette, France \label{paris}
\and Paris Region Fellow, Marie Sklodowska-Curie Action\label{paris_region_fellow}
\and Universidad Nacional Aut\'onoma de M\'exico, Instituto de Astronom\'ia, AP 106, Ensenada 22800, BC, M\'exico \label{ciudad1}
\and Cavendish Laboratory, JJ Thomson Avenue, Cambridge, CB3 0HE, UK \label{cavendish}
\and Instituto de Astrof\'isica de Andaluc\'ia (IAA-CSIC), Glorieta de la Astronom\'ia s/n, 18008 Granada, Spain \label{iaa}
\and Department of Astronomy and Astrophysics, University of California, Santa Cruz, 1156 High St. Santa Cruz, CA 95064, USA \label{ucsc}
\and Center for Astrophysics and Space Science, University of California San Diego, La Jolla, CA 92093, USA \label{ucsd}
\and Center for Interdisciplinary Exploration and Research in Astrophysics (CIERA), Northwestern University, 1800 Sherman, Evanston, IL 60201, USA \label{northwestern}
\and NASA Ames Research Center, Moffett Field, CA 94035, USA \label{nasa_ames}
\and Department of Physics, Massachusetts Institute of Technology, Cambridge, MA 02139, USA \label{DP_MIT}
\and Department of Aeronautics and Astronautics, MIT, 77 Massachusetts Avenue, Cambridge, MA 02139, USA \label{DAA_MIT}
\and Department of Physics and Kavli Institute for Astrophysics and Space Research, Massachusetts Institute of Technology, Cambridge, MA 02139, USA \label{DP&KIA&SR}
\and SETI Institute, Mountain View, CA  94043, USA \label{SETI}
\and Department of Astrophysical Sciences, Princeton University, Princeton, NJ 08544, USA \label{DAS_PU}
\and George Mason University, 4400 University Drive, Fairfax, VA 22030, USA \label{gmu} 
\and Komaba Institute for Science, The University of Tokyo, 3-8-1 Komaba, Meguro, Tokyo 153-8902, Japan \label{Komaba_Tokyo}
\and Instituto de Astrofisica de Canarias (IAC), 38205 La Laguna, Tenerife, Spain \label{iac_38205}
\and Department of Astronomy and Tsinghua Centre for Astrophysics, Tsinghua University, Beijing 100084, China \label{Tsinghua_china}
\and Astrobiology Center, 2-21-1 Osawa, Mitaka, Tokyo 181-8588, Japan \label{Osawa_Tokyo}
}

\titlerunning{\TESS\ discovery of a super-Earth orbiting the M-dwarf star TOI-1680 }\authorrunning{Ghachoui et al.}


\abstract{We report the discovery by the \TESS\ mission of a super-Earth on a 4.8-d orbit around an inactive M4.5 dwarf (\tar), validated by ground-based facilities. The host star is located 37.14 pc away, with a radius of $0.2100\pm0.0064$ \rsun, mass of $0.1800\pm0.0044$ \msun\,, and an effective temperature of 3211$\pm$100\,K. We validated and characterized the planet using \TESS\ data, ground-based multi-wavelength photometry from TRAPPIST, SPECULOOS, and LCO, as well as high-resolution AO observations from Keck/NIRC2 and Shane. Our analyses have determined the following parameters for the planet: a radius of $1.466^{+0.063}_{-0.049}$ \re\, and an equilibrium temperature of $404\pm14$\,K, assuming no albedo and perfect heat redistribution. Assuming a mass based on mass-radius relations, this planet is a promising target for atmospheric characterization with the James Webb Space Telescope (\JWST).
}


\maketitle
\section{Introduction} 

The science of exoplanets has dramatically flourished in the last decade, especially thanks to dedicated space missions. Following the completion of NASA's Kepler mission survey, where it was revealed that small transiting planets with sizes between those of Earth and Neptune (i.e., $1 < R_{p} < 4 \re$) are common in close-in orbits around other stars \citep{Howard_2012,Fressin2013}, the Transiting Exoplanet Survey Satellite \citep[TESS:][]{Ricker2015} mission took over to search for such planets orbiting bright and nearby stars \citep{Jenkins_2019}. This mission concept was chosen for easy subsequent spectroscopic investigation of the planets' masses and atmospheres, notably with the James Webb Space Telescope (\textit{JWST}) \citep{Deming2017ApJ}. \TESS\ observed 85\% of the sky in its nominal mission and is now in its extended mission \citep{Wong_2022BAAS}. Up to now, \TESS\ has detected more than 6000 planet candidates (TESS Objects of Interest, TOIs), including more than 1300 that could be smaller than 4 \re. 

The exploration of planets larger than Earth and smaller than Neptune is an area of great interest. Since such planets are not present in our Solar System, our understanding of their origins and formation mechanisms is limited. Interestingly, demographic studies performed by \citet{Fulton_2017} on the basis of California--Kepler Survey exoplanets sample --a subset of transiting planets from Kepler with high-resolution spectroscopic follow-up of their host stars \citep[CKS:][]{Petigura_2017,Johnson_2017}-- uncovered a gap, usually known as the “radius valley,” in the radius distribution of small planets in close orbits (<100 days) around FGK stars. This radius valley separates super-Earths and sub-Neptunes. This finding presents a key phenomenon for understanding planet formation mechanisms. 

Two main theories have been proposed to explain the radius valley: thermally driven mass-loss \citep{Lopez_2013, Owen_2013, Jin_2014, Chen_2016} and gas-poor formation  \citep{Luque_2021,Lee_2014, Lee_2016,Lee_2021}. Each of the two predict different origins for the radius valley. Moreover, a recent study conducted by \cite{Luque&palle2022} on a sample of 34 well-characterized exoplanets around M dwarfs has instead indicated the presence of a density gap separating rocky and water-rich exoplanets. However, the small size of exoplanet samples used in these studies precludes definitive constraints. Thus, having a significantly large sample of exoplanets with accurate density estimates is strongly needed. 

In this paper, we present the discovery and characterization of a super-Earth planet ($1.466^{+0.063}_{-0.049} \re$) which was first discovered by \TESS\ to orbit an M-dwarf star located near the continuous viewing zone (CVZ) of \JWST. We validate its planetary nature using ground-based observations, including time-series photometry, high-angular-resolution imaging and spectroscopy. Although we do not present a mass measurement in this paper, this could be done with high-precision radial velocity observation, as discussed in Sect.~\ref{discus}. This measurement would allow for detailed studies on planet formation in the future.

The paper is structured as follows. Section.~\ref{obs} presents the data from \TESS\ and all ground-based observations. Stellar characterization, validation of the transit signals, and transit analyses are presented in Sect.~\ref{anlys}. We discuss our findings in Sect.~\ref{discus} and give our conclusions in Sect.~\ref{concl}.

\section{Observations} \label{obs}

In this section, we present all the observations of \tar\ obtained with \TESS\ and ground-based facilities. Table \ref{tab:obs} summarizes all the ground-based, time-series photometric observations. 

\begin{table*}
\begin{center}
\caption{Ground-based time-series photometric observations of \tar, and the detrending parameters that maximize the log-likelihood of each light curve. \label{tab:obs}}
\begin{tabular}{l c c c c c}
\toprule
Date (UT) & Filter &  Facility & Exp. time [s] & Notes & Detrending parameters\\
\midrule
13  Jun 2020 & $Ic$          & LCO-McD-1m         & 150  & full           & Airmass              \\
22  Jun 2021 & $z'$          & TRAPPIST-N-0.6m    & 80   & egress         & Airmass              \\
07  Jul 2020 & $i'$          & LCO-McD-1m         & 150  & full           & Airmass              \\
16  Jul 2021 & $z'$          & TRAPPIST-N-0.6m    & 80   & full           & Airmass + Background \\
09 Aug 2021  & $r'$          & Artemis-1m         & 100  & full           & Airmass              \\
02 Sep 2021  & $I+z$         & Artemis-1m         & 20   & full           & Airmass + Background \\
21 Sep 2021  & $I+z$         & SAINT-EX-1m        & 19   & full           & Airmass + Background \\
08 Nov 2021  & $r'$          & SAINT-EX-1m        & 105  & No post-egress & Airmass              \\
21 Apr 2022  & $g',r',i',zs$ & LCO-Hal-M3-2m      & 200,116,58,56  & full & Airmass              \\
\hline 
\end{tabular}
\end{center}
\end{table*}

\subsection{\TESS\ photometry} \label{TESS:phot}
\begin{figure*}

\centering

\includegraphics[width=0.4\textwidth]{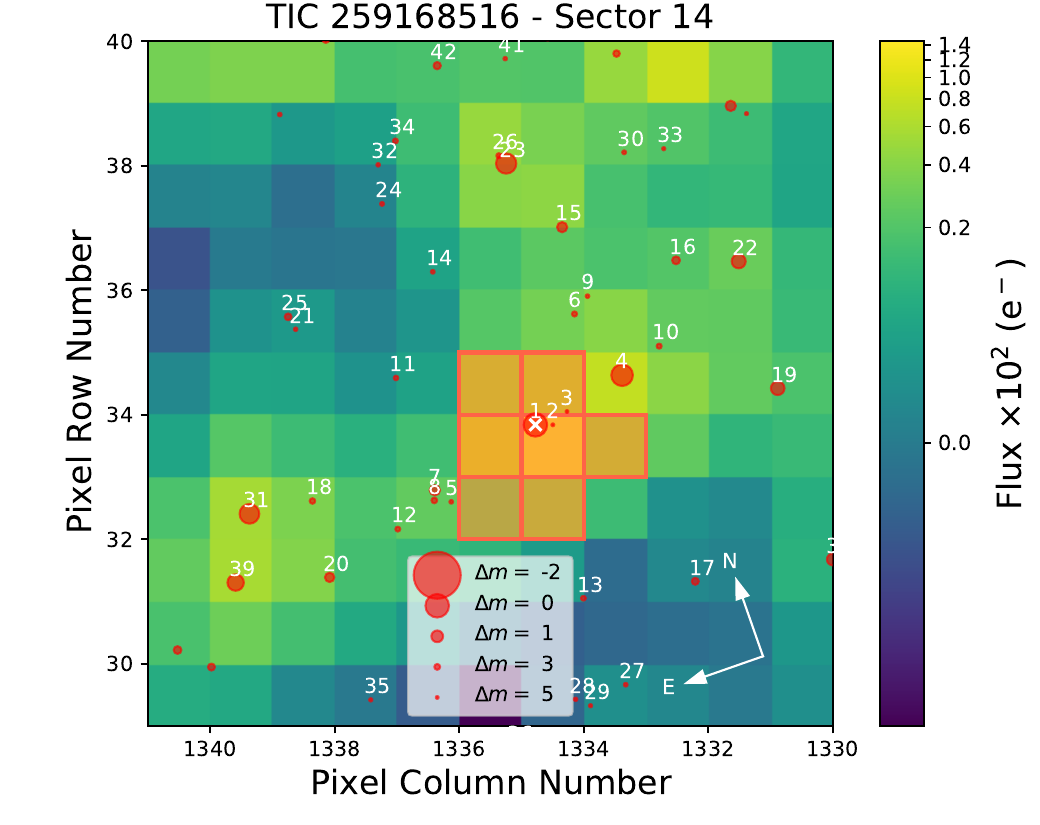}
\includegraphics[width=0.4\textwidth]{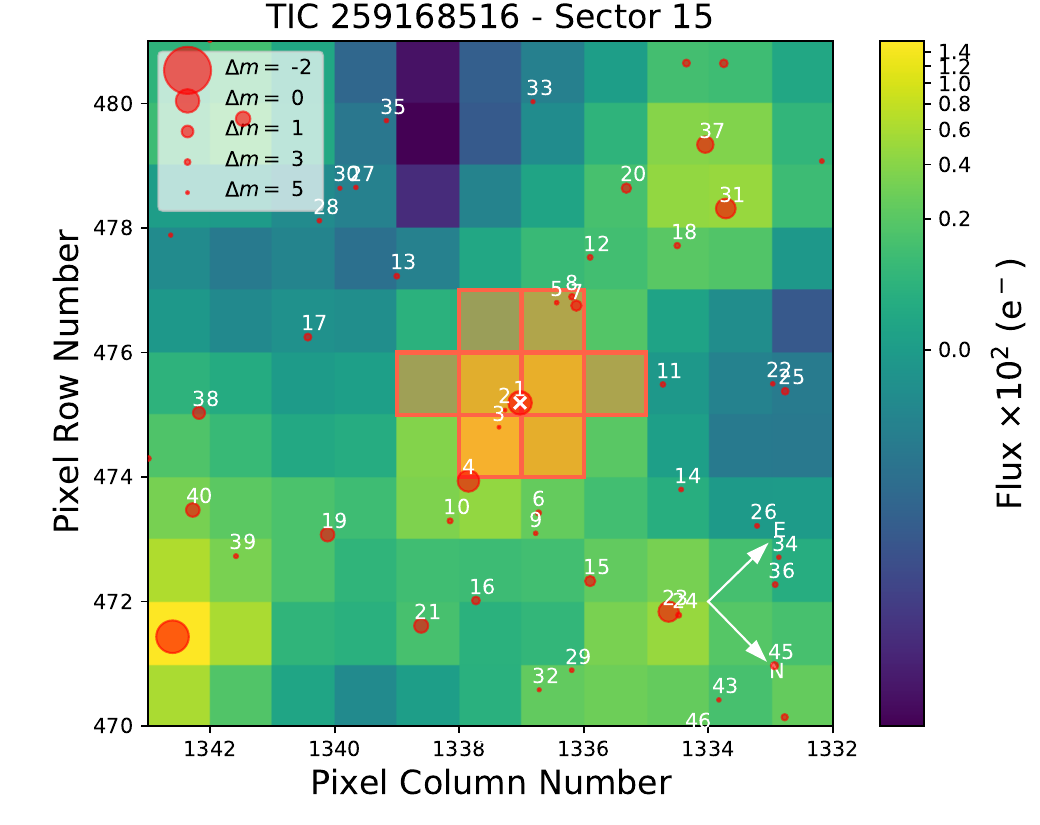}
\caption{Target pixel files \textcolor{red}{}(TPFs) of \tar\ in \TESS\ Sector 14 and 15, created with \texttt{tpfplotter} \citep{Aller_2020A&A}. The orange shaded region represents the aperture used to extract the SPOC photometry. The red circles are the \gaia\ DR2 sources. Sizes represent magnitude contrasts with respect to \tar. Figure continued in Appendix \ref{appendix}.} 
\label{fov}
\end{figure*}

Over its two-year primary mission, \TESS\ \citep{Ricker2015} performed an all-sky survey in a series of contiguous overlapping $96 \times 24~\rm deg$ sectors, each observed for 27 days. Depending on the ecliptic latitude, the overlapping regions of the sectors were observed for up to $\sim$351 days. Given its high ecliptic latitude ($\beta = +81.05$ deg), \tar\ (TIC 259168516) is well placed in the \TESS\ CVZ. It was then observed by \TESS\ in  all the northern sectors (from 14 to 26) in the second year of \TESS\ primary mission, from 18 July 2019 to 4 July 2020. It was also observed in the \TESS\ extended mission in sectors 40-41 from 25 June to 20 August 2021. Most recently, it was observed in sectors 47--59 from 31 December 2021 to 23 December 2022. The target pixel files (TPFs) and simple aperture photometry (SAP) apertures used in each sector are shown in Fig.~\ref{fov}, along with the superplotted locations of nearby Gaia DR2 \citep{gaia_2018} sources. The astrometric and photometric properties of \tar\ from the literature are reported in Table \ref{tableTOI-1680}. The time series observations were processed in the \TESS\ Science Processing Operations Center (SPOC) pipeline, originally developed for the Kepler mission at NASA Ames Research Center \citep{Jenkins_2016,Jenkins_2020a}. The SPOC pipeline conducted a transit search of the combined light curve from sectors 14-16 on 26 October 2019 with an adaptive, noise-compensating matched filter \citep{Jenkins_2002,Jenkins_2010,Jenkins_2020b}, producing a threshold crossing event (TCE) with 4.8 day period for which an initial limb-darkened transit model was fitted \citep{LI_2019PASP} and a suite of diagnostic tests were conducted to help make or break the planetary nature of the signal \citep{Twicken_2018PASP}. The 5.1 ppt transit signature passed all diagnostic tests presented in the SPOC data validation reports, and the source of the transit signal was localized within $4.03\pm4.58$ \arcsec. The TESS Science Office (TSO) reviewed the vetting information and issued an alert for \tar\,b on 30 January 2020 \citep{Guerrero_2021ApJ}. 

For subsequent analysis, we retrieved the 2-minute presearch data conditioning light curves (PDC-SAP, \citealt{Stumpe_2012,Stumpe_2014,Smith_2012}) from the Mikulski Archive for Space Telescopes (MAST). We were limited to sectors 13--26 of the primary mission and sectors 47--50 of the extended mission. We removed all the bad data points flagged as "bad quality." We then detrended the light curves to remove stellar variability using a biweight time-windowed slider via \texttt{wotan} \citep{Hippke2019}. We excluded the transit signal by applying a filter window that is three times longer than the transit duration of $71.150_{-0.936}^{+0.993}$ minutes.

\subsection{Ground-based photometry}\label{gbp}

The pixel scale of \TESS\ spacecraft is 21\arcsec\ per pixel \citep{Ricker2015}. A targeted star might not be alone in a single pixel. Other stars in the same pixel might be suspected to be the source of the \TESS\ detection. Even if the transit signal is on target, the depth might appear shallower because of the contaminating nearby stars. To confirm the signal on target and validate its planetary nature, a series of precise ground-based observations were collected using five observatories as part of the \TESS\ Follow-up Observing Program 
(TFOP\footnote{\url{https://tess.mit.edu/followup}}). We made use of \TESS\ Transit Finder (\texttt{TTF}) tool, which is a customized version of the \texttt{Tapir} software package \citep{Jensen2013}, to schedule our observations described hereafter.

\subsubsection{LCOGT 1m}

The first two full transits of \tar\,b were observed from Las Cumbres Observatory Global Telescope (LCOGT; \citealt{Brown2013}) 1.0-meter network node at McDonald Observatory. The 1-meter telescopes are equipped with 4096$\times$4096 pixels SINISTRO cameras having a pixel scale of 0.389\arcsec\ per pixel, offering a field of view of $26'{\times}26'$. The first transit was observed on 13 June 2020 in the \textit{Ic} band for 210 min, over which we gathered 64 images with an exposure time of 150 seconds. The second transit was observed on 07 July 2020 in Sloan $i'$ band during an observational window of 198 min, where we collected 63 images with an exposure time of 150 seconds. The data reduction and photometric data extraction were performed using the \texttt{AstroImageJ} \citep[AIJ:][]{Karen2017} software package with an uncontaminated aperture of 8.0 pixels (3.11\arcsec) for both observations.

\subsubsection{TRAPPIST-North photometry}

We observed a partial and a full transit of \tar\,b with the 0.6-meter TRAPPIST-North telescope located at Oukaimeden Observatory in Morocco (\citealt{jehin2011,gillon2013,Barkaoui2019}) on 22 June and 16 July 2021, respectively. TRAPPIST-North is equipped with a thermoelectrically cooled 2K${\times}$2K Andor iKon-L BEX2-DD CCD camera with a pixel scale of 0.6\arcsec per pixel, offering a field of view of $20'{\times}20'$. Both observations were performed in Sloan $z'$ band with an exposure time of 80\,s. The first observation consisted of 113 images for 182 min and the second consisted of 99 images for 166 min duration. For both datasets, we performed the data reduction and differential aperture photometry using \texttt{prose}\footnote{\url{https://github.com/lgrcia/prose}} \citep{2022MNRAS.509.4817G}, which selected the optimum apertures for the photometric data extraction to be 6.94 pixels (4.16\arcsec) for the first observation and 8.32 pixels (5\arcsec) for the second.


\subsubsection{SPECULOOS-North Artemis photometry}

Two full transits of \tar\,b were observed by the  telescope Artemis of the SPECULOOS Northern Observatory (SNO, \citealt{SNO_2022}), located at the Teide Observatory (Canary Islands, Spain). Artemis is operated in a fully automated manner and equipped with Andor iKon-L camera with a 2K$\times$2K deep-depletion CCD, which has a pixel scale of 0.35\arcsec\ per pixel. The first transit was observed on 09 August 2021 in the Sloan $r'$ filter with an exposure time of 100\,s. We gathered 156 images over 319\,minutes. The second transit was observed on 02 September 2021 in an $I+z$ filter (Johnson/Cousins $I$ + Sloan $z'$) with an exposure time of 20\,s. We gathered 641 images during an observational window of 325 minutes. Both datasets were calibrated, and the differential aperture photometry were performed using the \texttt{PRINCE} pipeline \citep{Demory2020}. The aperture radii used were 5.0 pixels (1.75\arcsec) for the first observation and 8.5 pixels (2.97\arcsec) for the second.

\subsubsection{SAINT-EX photometry}

More ground-based photometric time-series observations of \tar\,b were obtained from the SAINT-EX observatory (Search And characterIsatioN of Transiting EXoplanets). SAINT-EX is a 1-m telescope in the F/8 Ritchey-Chr\'etien configuration and operated in fully robotic manner. It is equipped with an 2k$\times$2k deep-depletion CCD camera with a pixel scale of 0.34\arcsec\ per pixel offering a field of view of $12'{\times}12'$. The telescope is allied to an ASTELCO equatorial NTM-1000 German mount associated with direct-drive motors that permits observation without a meridian flip. It is in fact a twin of the SPECULOOS-South and SPECULOOS-North telescopes, and it operates as part of the SPECULOOS survey \citep{Delrez_2018SPIE,Sebastian2021A&A}.  

Two full transits were observed by SAINT-EX. The first on 21 September 2021 in an $I+z$ filter for 314 min, in which we gathered 520 raw images with an exposure time of 19\,s. And the second on 08 November 2021 in Sloan $r'$ filter for 206 min, with an exposure time of 105 seconds where we gathered 104 images. The data reduction and differential aperture photometry were performed automatically using the \texttt{PRINCE} pipeline. For more information on the SAINT-EX telescope and \texttt{PRINCE} pipeline, we refer to \cite{Demory2020}.
The aperture radii used were 6.5 pixels (2.28\arcsec) for the first observation and 11.0 pixels (3.85\arcsec) for the second.

\subsubsection{LCOGT MUSCAT3 photometry}

A full transit of \tar\,b was observed simultaneously in Sloan-$g'$, $r'$, $i'$, and Pan-STARRS $z$-short bands on UTC April 21, 2022 using the LCOGT 2\,m Faulkes Telescope North at Haleakala Observatory on Maui, Hawaii. The telescope is equipped with the MuSCAT3 multi-band imager \citep{Narita_2020SPIE11447E}. The raw images were calibrated using the standard LCOGT {\tt BANZAI} pipeline \citep{McCully_2018SPIE10707E}, and photometric measurements were extracted using {\tt AstroImageJ} \citep{Karen2017}. The light curve in the Sloan-$g'$ filter was not selected to be included in the analyses because of the low signal-to noise ratio (S/N) that is due to the faintness of the star in this band.

\subsection{Spectroscopic observations}

With the aim of better constraining the stellar properties, we also performed spectroscopic observation detailed hereafter. The analyses are presented in Sect.~\ref{sec:star_spec}.

\subsubsection{IRTF/SpeX}

We gathered a near-infrared spectrum of \tar\ with the SpeX spectrograph \citep{Rayner2003} on the 3.2-m NASA Infrared Telescope Facility (IRTF) on 19 Oct 2021 (UT). The conditions were clear with a seeing of 1$\farcs$0--1$\farcs$2.
We followed the same observational design as other recent IRTF/SpeX observations of M-dwarf TOIs \citep{Wells2021, Delrez2022}. We used the short-wavelength cross-dispersed (SXD) mode with the $0.3\arcsec\times 15\arcsec$ slit aligned to the parallactic angle, which gives a set of spectra covering 0.75--2.42\,$\mu$m with a resolving power of $R{\sim}2000$. Nodding in an ABBA pattern, we collected 18 exposures of 64.9\,s each, totaling 19.5\,min on source. We collected a set of standard SXD flat-field and arc-lamp exposures immediately after the science frames, followed by a set of six, 2.8-s exposures of the A0\,V star HD\,172728 ($V{=}5.7$). We reduced the data using Spextool v4.1 \citep{Cushing2004}, following the instructions for standard usage in the Spextool User's Manual\footnote{Available at \url{http://irtfweb.ifa.hawaii.edu/~spex/observer/}}. The final spectrum has a median S/N per pixel of 68 with peaks in the $J$, $H$, and $K$ bands of 98, 101, and 91, respectively, along with an average of 2.5 pixels per resolution element.

\subsubsection{Shane/Kast}
We obtained a low-resolution optical spectrum of \tar\ on 27 Nov 2021 (UT) using the Kast double spectrograph \citep{kastspectrograph} on the 3-m Shane Telescope at Lick Observatory. Conditions were partly cloudy with a seeing of 1$\arcsec$. We obtained two sequential exposures of 1200\,s (40 minutes total) through the red channel of Kast using the 600/7500 grism and 2$\arcsec$-wide slit, providing spectra covering 5900--9200~{\AA} at an average resolving power of $R \approx$ 1900. We observed the spectrophotometric calibrator Feige 110 \citep{1992PASP..104..533H,1994PASP..106..566H} earlier that night for flux calibration, and the G2~V star HD~205113 (V = 6.87) immediately after \tar\ for telluric absorption calibration. Flat-field and arc line lamps were obtained at the start of the night for flux and wavelength calibration. Data were reduced and analyzed using the \texttt{kastredux} package\footnote{\url{https://github.com/aburgasser/kastredux}}, with standard settings for image reduction and calibration, boxcar extraction of the spectrum, wavelength calibration, flux calibration, and telluric absorption calibration. The final spectrum has a S/N = 150 at 7500~{\AA} and wavelength accuracy of 0.51~{\AA} (22~km/s).

\subsection{High-Resolution Imaging}

As part of our standard process for validating transiting exoplanets to assess the possible contamination of bound or unbound companions on the derived planetary radii \citep{ciardi2015}, we observed the \tar\ with near-infrared adaptive optics (AO) imaging at Keck and Shane Observatories. \gaia\ DR3 is also used to provide additional constraints on the presence of undetected stellar companions as well as wide companions.
    
\subsubsection{Keck-II near-infrared adaptive optics omaging} \label{NIRC2}

The Keck Observatory observations were made with the NIRC2 instrument on Keck-II behind the natural guide star AO system \citep{wizinowich2000} on 28 Aug 2021 UT in the standard three-point dither pattern that is used with NIRC2 to avoid the left lower quadrant of the detector, which is typically noisier than the other three quadrants. The dither pattern step size was $3\arcsec$ and was repeated twice, with each dither offset from the previous dither by $0.5\arcsec$.  NIRC2 was used in the narrow-angle mode with a full field of view of $\sim10\arcsec$ and a pixel scale of approximately 0.0099442\arcsec\ per pixel.  The Keck observations were made in the $K$ filter $(\lambda_o = 2.196; \Delta\lambda = 0.336~\mu$m) with an integration time of 1 second for a total of 9 seconds on target.
    
The AO data were processed and analyzed with a custom set of IDL tools.  The science frames were flat-fielded and sky-subtracted.  The flat fields were generated from a median average of dark-subtracted flats taken on-sky. The flats were normalized such that the median value of the flats is unity. The sky frames were generated from the median average of the dithered science frames; each science image was then sky-subtracted and flat-fielded.  The reduced science frames were combined into a single combined image using an intra-pixel interpolation that conserves flux, shifts the individual dithered frames by the appropriate fractional pixels, and median-coadds the frames.  The final resolutions of the combined dithers were determined from the full-width half-maximum of the point spread functions (0.056\arcsec\ for the Keck observations).  
	
\begin{figure}
    \centering
    \includegraphics[width=0.5\textwidth]{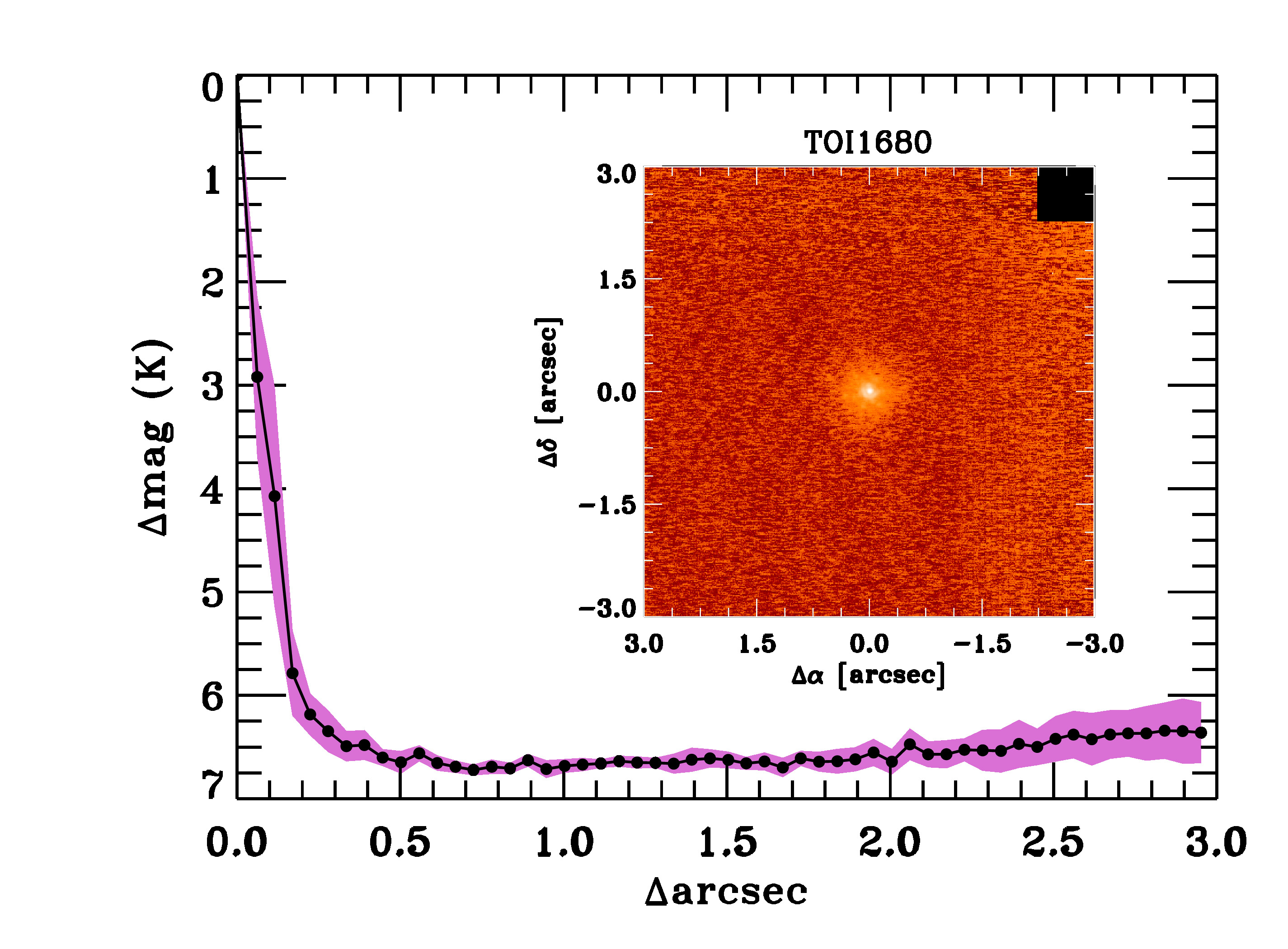}
    \caption{Keck NIR AO imaging and sensitivity curve for \tar\ taken in the $K$ filter. The image reaches a contrast of ${\sim}7$ magnitudes fainter than the host star beyond 0.\arcsec5. {\it Inset:} Image of the central portion of the data.
    }
    \label{fig:ao_images}
\end{figure}

The sensitivities of the final combined AO image were determined by injecting simulated sources azimuthally around the primary target every $20^\circ $ at separations of integer multiples of the central source's FWHM \citep{furlan2017}. The brightness of each injected source was scaled until standard aperture photometry detected it with 
$5\sigma $ significance. The resulting brightness of the injected sources relative to \tar\ set the contrast limits at that injection location. The final $5\sigma $ limit
at each separation was determined from the average of all of the determined limits at that separation and the uncertainty on the limit was set by the rms dispersion of the azimuthal slices at a given radial distance.  The Keck data have a sensitivity close-in of $\delta \mathrm{mag} = 2.9$ mag at 0.06\arcsec, and deeper sensitivity at wider separations ($\delta \mathrm{mag} = 6.5$ mag at $\gtrsim$0.4\arcsec). The final sensitivity curve for the Keck is shown in Fig~\ref{fig:ao_images}.  No close-in ($\lesssim 1\arcsec$) stellar companions were detected by Keck.   
    
\subsubsection{Shane near-infrared adaptive optics imaging}

\begin{figure*}[!h]
    \centering
    \includegraphics[width=0.45\textwidth]{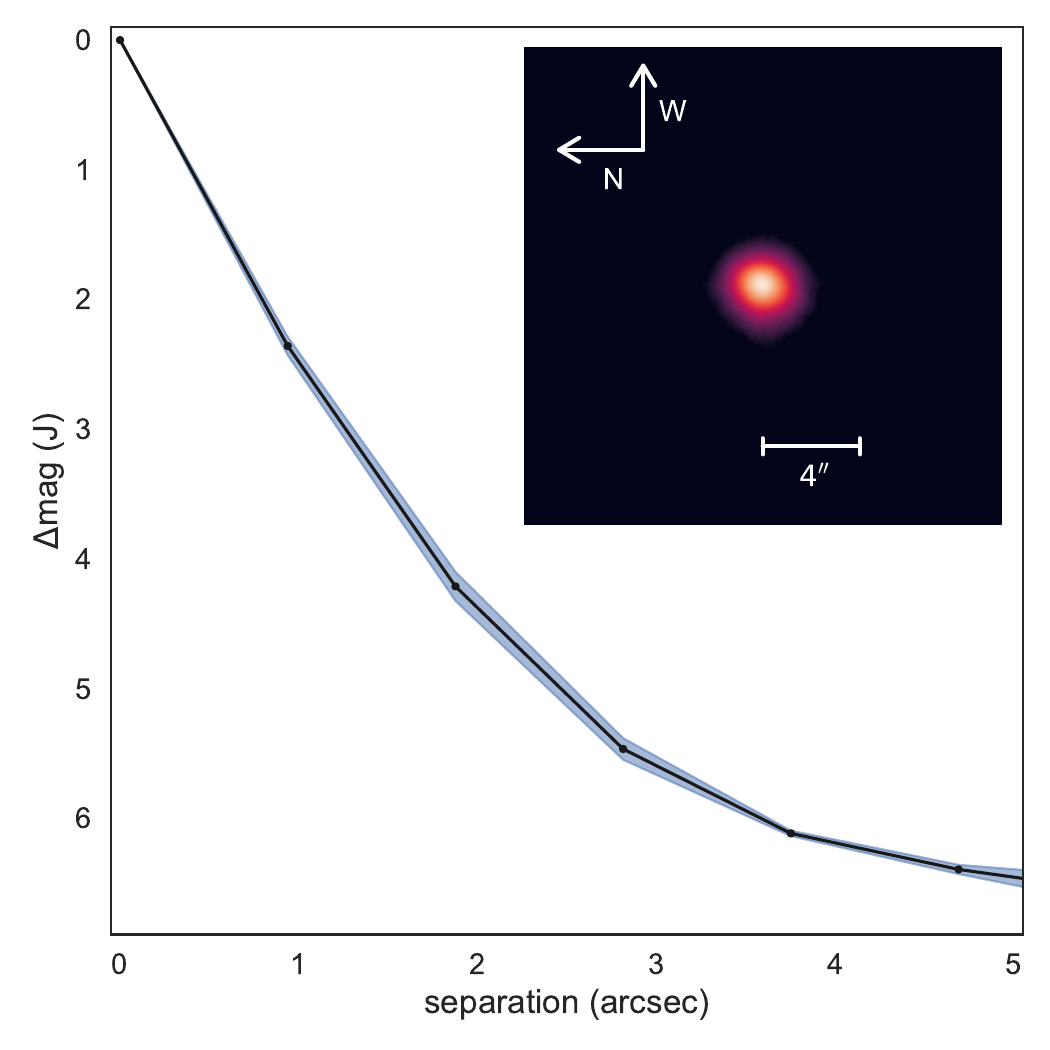}
    \includegraphics[width=0.45\textwidth]{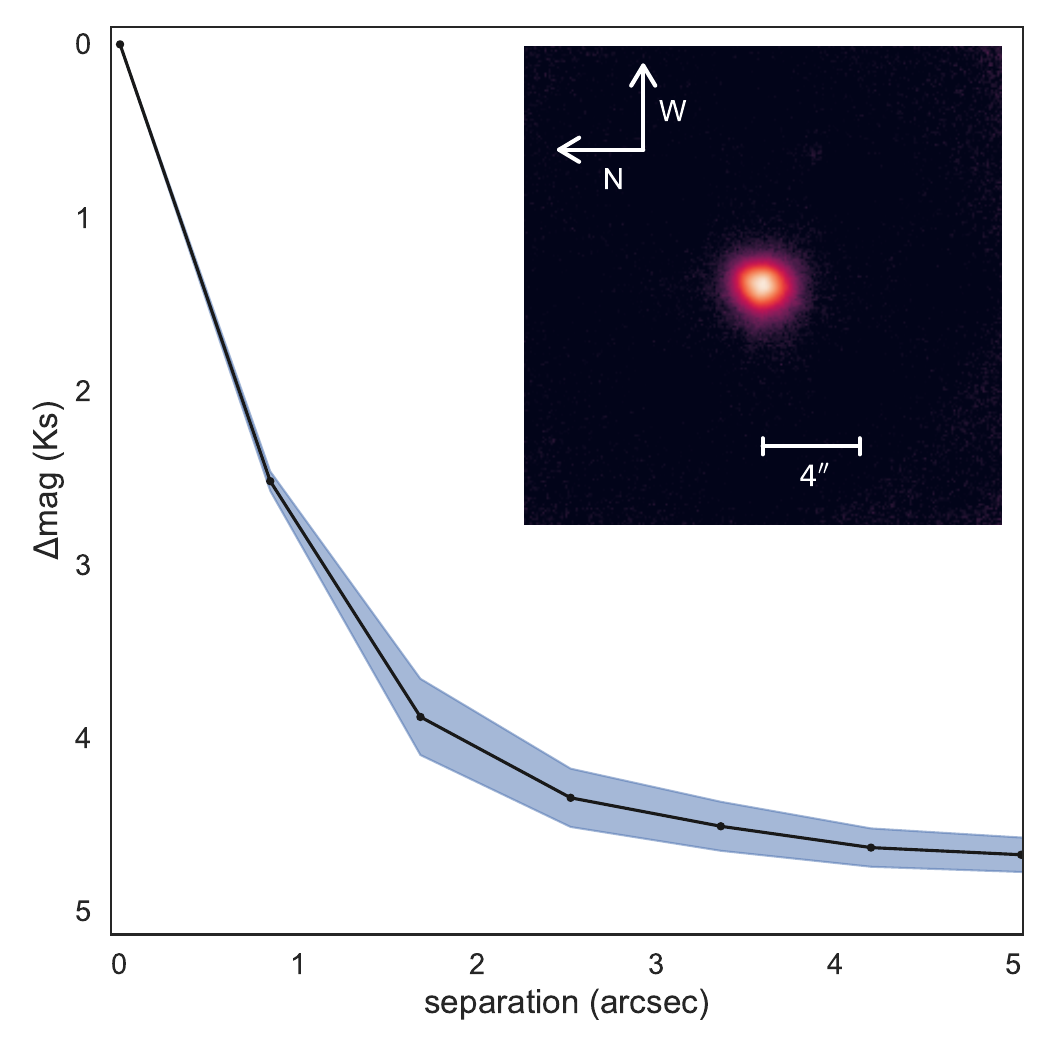}
    \caption{High-resolution imaging and contrast curves of TOI-1680 obtained with the ShARCS camera on the Shane 3-m telescope, with the $J$ (left) and $Ks$ filter (right). No secondary sources were detected.}
    \label{fig:TOI1680_sharcs_J}
\end{figure*}

We observed TIC 259168516 on UT 2021 June 2 using the ShARCS camera on the Shane 3-meter telescope at Lick Observatory \citep{2012SPIE.8447E..3GK, 2014SPIE.9148E..05G, 2014SPIE.9148E..3AM}. The observation was taken with the Shane adaptive optics system in natural guide star mode. The final images were constructed using sequences of images taken in a four-point dither pattern with a separation of 4$\arcsec$ between each dither position. Two image sequences were taken of this star: one with a $Ks$ filter ($\lambda_0 = 2.150$ $\mu$m, $\Delta \lambda = 0.320$ $\mu$m) and one with a $J$ filter ($\lambda_0 = 1.238$ $\mu$m, $\Delta \lambda = 0.271$ $\mu$m). A more detailed description of the observing strategy and reduction procedure can be found in \cite{2020AJ....160..287S}. The contrast curves extracted from these observations are shown in Fig~\ref{fig:TOI1680_sharcs_J}. With the $Ks$ filter, we achieve contrasts of 2.5 at $1 \arcsec$ and 4.4 at $2 \arcsec$. With the $J$ filter, we achieve contrasts of 2.8 at $1 \arcsec$ and 4.0 at $2 \arcsec$. We detect one companion about $5\farcs8$ west of TIC 259168516 that is 5.0 magnitudes fainter in $Ks$ and 5.7 magnitudes fainter in $J$. Based on this, the star is likely the known neighbor TIC 1884271108. Gaia EDR3 parallax and proper motion indicate that it is another line-of-sight star.

\subsection{\gaia\ assessment}
In addition to the high resolution imaging, we have used \gaia\ to identify any wide stellar companions that may be bound members of the system.  Typically, these stars are already in the \TESS\ Input Catalog and their flux dilution to the transit has already been accounted for in the transit fits and associated derived parameters from the TESS PDC-SAP photometry.  There are no additional widely separated companions identified by \gaia\ that have the same distance and proper motion as \tar\ \citep[see also ][]{mugrauer2020,mugrauer2021}.
     
Additionally, the \gaia\ DR3 astrometry provides additional information on the possibility of inner companions that may have gone undetected by either \gaia\ or the high resolution imaging. The \gaia\ renormalised unit weight Error (RUWE) is a metric, similar to a reduced chi-square, where values that are $\lesssim 1.4$  indicate that the \gaia\ astrometric solution is consistent with the star being single whereas RUWE values $\gtrsim 1.4$ may indicate an astrometric excess noise, possibly caused by the presence of an unseen massive (stellar) companion \citep[e.g., ][]{ziegler2020}.  \tar\ has a \gaia\ DR3 RUWE value of 1.05 indicating that the astrometric fits are consistent with the single star model.

\section{Analyses} \label{anlys}

\subsection{Stellar characterization} \label{star}

\subsubsection{Spectroscopic analysis}\label{sec:star_spec}

\begin{figure}
    \centering
    \includegraphics[width=\columnwidth]{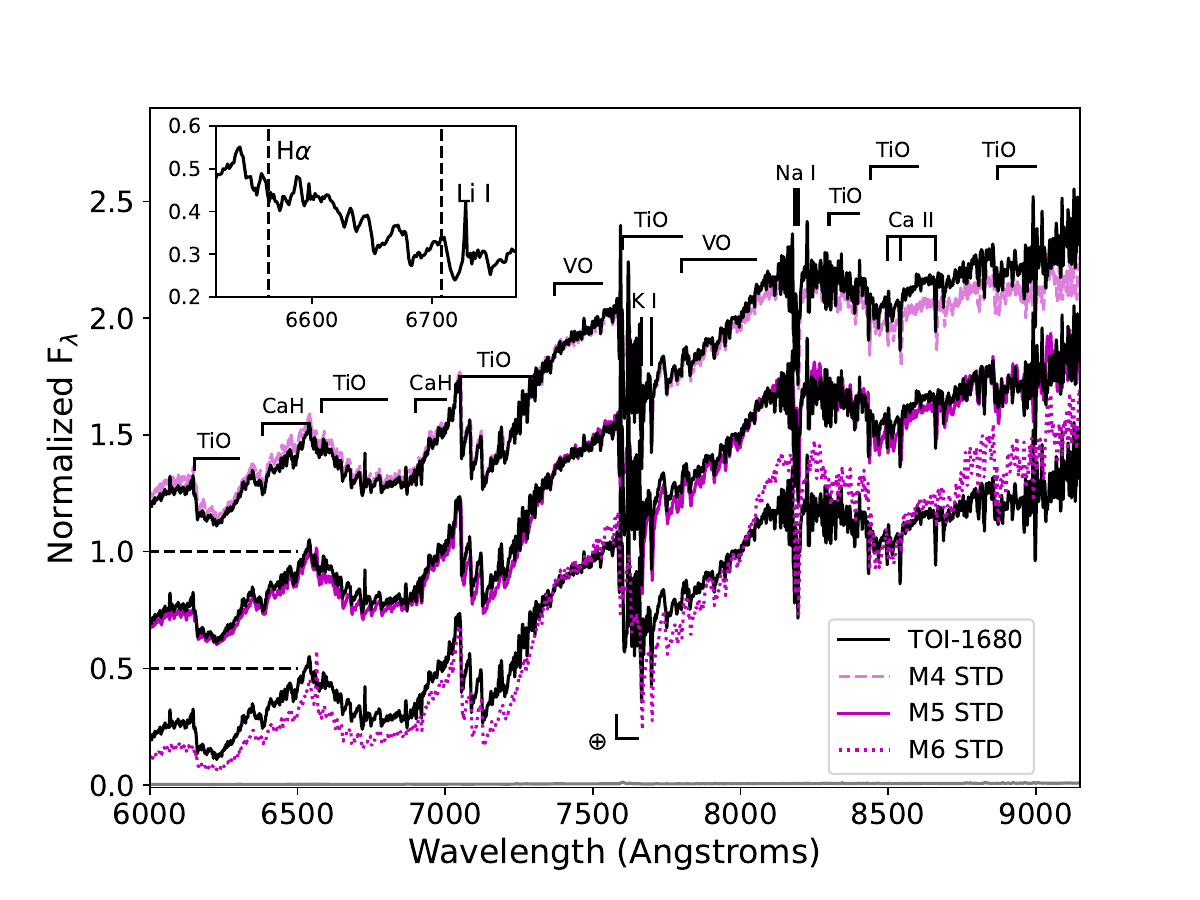}
    \includegraphics[width=\columnwidth]{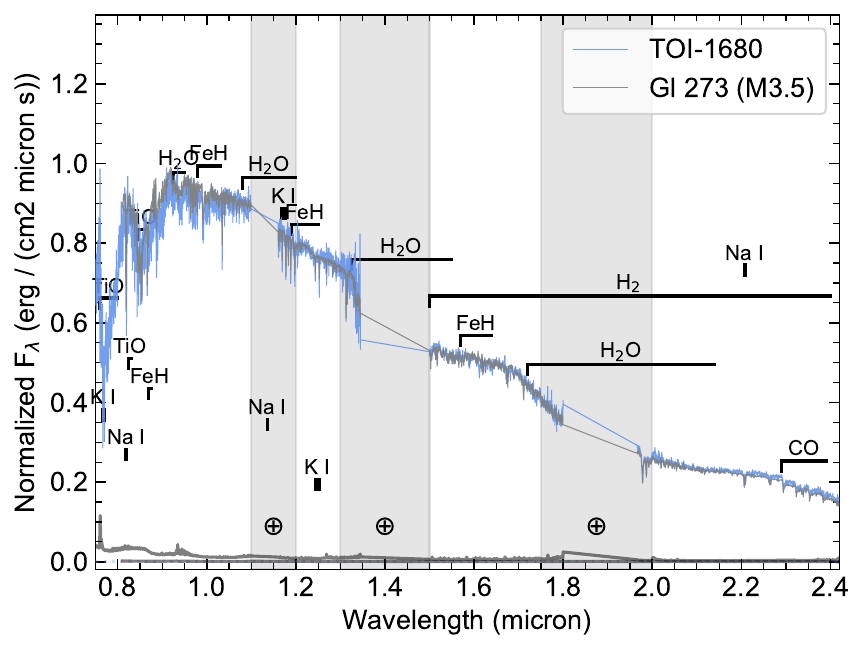}
    \caption{
    Kast optical spectrum of \tar\ (black lines) compared to M4, M5, and M6 SDSS optical templates (magenta lines from top to bottom) from \citet{2007AJ....133..531B} shown at the top. Each spectral comparison is normalized at 7500~{\AA} and offset by a constant of 0.5. Key atomic and molecular absorption features are labeled, including residual O$_2$ telluric absorption at 7600~{\AA} ($\oplus$ symbol). The inset box shows a close-up of the region around the 6563~{\AA} H$\alpha$ and 6708~{\AA} Li~I lines, neither of which are detected. Bottom panel displays SpeX SXD spectrum of \tar\ (blue) compared to SpeX SXD spectrum of the M3.5 standard Gl\,273 \citep[grey;][]{Rayner2009}. Prominent spectral features of M dwarfs are highlighted, and regions of strong telluric absorption are shaded. The grey lines at the bottom of the plot show the measurement uncertainties for \tar\ and Gl\,273 (higher and lower, respectively).
    }
    \label{fig:spectracomp}
\end{figure}

The Shane/Kast optical and IRTF/SpeX near-infrared spectra allow us to assess \tar's fundamental stellar properties.
Using tools in the \texttt{kastredux} package, we compared the optical spectrum to the SDSS M dwarf templates of \citet{2007AJ....133..531B}, and found a best overall match to the M5 template (Fig.~\ref{fig:spectracomp}). Spectral indices from \citet{1995AJ....110.1838R}, \citet{1999AJ....118.2466M}, \cite{2003AJ....125.1598L}, and \citet{2007MNRAS.381.1067R} are more consistent with an M4 classification, suggesting an intermediate type of M4.5$\pm$0.5.
The $\zeta$ metallicity index of \citet{2007ApJ...669.1235L,2013AJ....145...52M}, based on relative strengths of TiO and CaH features, is measured to be 1.025$\pm$0.002, consistent with a metallicity of $\mathrm{[Fe/H]}$ = $+$0.04$\pm$0.20 based on the calibration of \citet{2013AJ....145...52M}.
We see no evidence of H$\alpha$ emission in the Balmer line at 6563~{\AA} (equivalent width limit of $<$0.3~{\AA}), suggesting an age $\gtrsim$4--7~Gyr \citep{2008AJ....135..785W}.

The SpeX SXD spectrum of \tar\ is shown in Fig.~\ref{fig:spectracomp}.
We used the SpeX Prism Library Analysis Toolkit \citep[SPLAT, ][]{splat} to compare the spectrum to that of single-star spectral standards in the IRTF Spectral Library \citep{Cushing2005, Rayner2009}, finding the best single match to the M3.5 standard Gl\,273. We note that the shape of the spectrum of \tar\ suggests it is cooler than the M3.5 standard, though the M4.0 standards in the library give poorer fits.
We adopt an infrared spectral type of M3.5 $\pm$ 0.5, earlier but consistent with the optical classification. After adjusting for a barycentric velocity of $-1.64$\,km/s, we cross-correlated the SpeX spectrum of \tar\ with the rest-frame velocity of the M3.5 standard to determine the radial velocity. Determining the uncertainty of the cross-correlation with a Monte Carlo approach, we estimate a radial velocity of $-34.3 \pm 3.3$\,km/s. After applying a radial-velocity correction, we confirmed that the best-fit spectral standard did not change.

The SpeX spectrum also provides an estimate of stellar metallicity.
Using SPLAT, we measured the equivalent widths of the $K$-band Na\,\textsc{i} and Ca\,\textsc{i} doublets and the H2O--K2 index \citep{Rojas-Ayala2012}.
We then used the \citet{2013AJ....145...52M} relation between these observables and [Fe/H] to estimate the stellar metallicity, propagating uncertainties using a Monte Carlo approach (see \citealt{Delrez2022}).
We determined a metallicity of $\mathrm{[Fe/H]} = -0.32 \pm 0.13$, which is lower than but formally consistent with the optical measurement and more in line with the apparent old magnetic activity age of the star.

\subsubsection{Empirical relations} \label{em_rel}

We used available empirical relationships appropriate for M dwarfs to determine the stellar parameters of \tar. We first used \gaia\ EDR3 parallax and 2MASS $m_{K}$ visual magnitude to calculate the $M_{K}$ absolute magnitude and found $M_{K} = 7.9720 \pm 0.0203$. We then used the empirical relationship between the mass and $M_{K}$ absolute magnitude of \cite{Mann_2019} to estimate the mass of \tar, which we found to be $M_\star = 0.1800 \pm 0.0043 $ \msun. This is in good agreement with the mass value of $M _\star = 0.1765 \pm 0.02$ \msun\,, estimated using the mass--luminosity relation in the K-band from \cite{benedict_2016}, where the uncertainty is dominated by the scatter in the mass-Ks relation.

Using the empirical polynomial relation between the stellar radius $R_{*}$ and $M_{K}$ derived by \cite{Mann_2015}, we estimated $R_{*}$ to be $0.2130 \pm 0.0064 $ \rsun, with a typical uncertainty of 3\% as reported in Table 1 of \cite{Mann_2015}. As an independent check, we used the mass--radius relationship of \cite{Boyajian_2012} to determine the radius from the masses we previously estimated. We found 
$ R_{*}=0.2075 \pm 0.0039$ \rsun, which is consistent with the aforementioned radius determination. This leads to a stellar density of $27.4\pm2.6~\rm \,\mathrm{g\,cm^{-3}}$.

As for the effective temperature determination, we first estimated the bolometric correction in the K-band  to be  $BC_{K} = 2.7414 \pm 0.0822 $ mag, by making use of the empirical polynomial relation between $BC_{K}$ and $V-J$ of \cite{Mann_2015}. Then, we determined a bolometric magnitude of $M_{bol} = 10.72 \pm 0.085$ mag, which gives a bolometric luminosity of $ L_{*} = 0.004353 \pm 0.000340 \lsun$. The Stefan-Boltzmann Law, along with the aforementioned stellar radius and bolometric luminosity, gives an effective temperature $T_{\rm eff} = 3210 \pm 62$\,K. Independently, we also determined the effective temperature based on the empirical relation of \cite{Mann_2015} using the color indexes $V-J$ and $J-H$, and found $T_{\rm eff} = 3224 \pm 100$\,K. The two values are consistent within 1$\sigma$.

\subsubsection{SED fitting}

As an independent check, we used the \texttt{EXOFASTv2} analyses package \citep{Eastman_2019} to perform an analysis of the broadband spectral energy distribution (SED) of \tar\ using MIST stellar models \citep{Dotter_2016,Choi_2016} to determine the stellar parameters. We pulled the \textit{$JHK_{s}$} magnitudes from the 2MASS catalog \citep{Cutri_2003}, WISE1-WISE4 magnitudes from the AllWISE catalog \citep{Cutri_2003} and the $G G_{BP} G_{RP}$ magnitudes from \gaia\ EDR3 (see Table \ref{tableTOI-1680}). We performed the fit with $T_{\rm eff}$, $R_{*}$ and $M_{*}$ as free parameters with a Gaussian prior on the \gaia\ EDR3 parallax which we corrected for systematics by subtracting -0.041867248 mas from the nominal value according to the \cite{Lindegren_2020} prescription.
We set an upper limit on the extinction of $A_{V} = 0.29233$ from the dust maps of \cite{Schlafly&Finkbeiner2011} and a Gaussian prior on the stellar metallicity from  IRTF/SpeX (see Table \ref{table:star}). The SED fit results, reported in Table \ref{table:star}, are in excellent agreement with our previous determinations. 

\begin{figure}
	\includegraphics[width=0.51\textwidth]{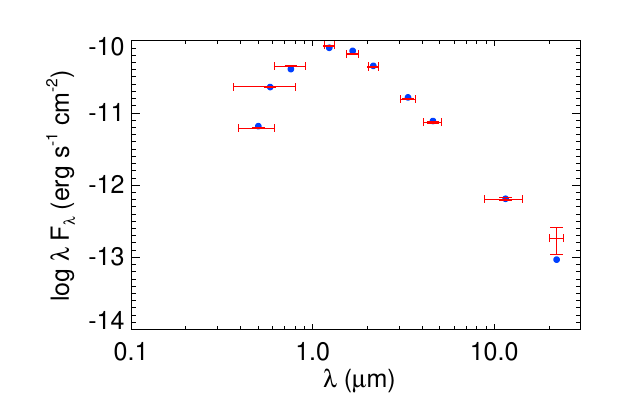}
	\caption{Spectral energy distribution (SED) of \tar. The red crosses show the broadband observations and the error bars show the width of the filters. The blue circles represent the model fluxes.}

	\label{fig:sed}
\end{figure}

\begingroup 
\begin{table} 
	\caption{
		\tar\ stellar astrometric and photometric properties.} 
	\begin{center}
		\renewcommand{\arraystretch}{1.15}
		\begin{tabular*}{\linewidth}{@{\extracolsep{\fill}}l c c}
			\toprule
			Parameter & Value & Source \\
			\midrule
			\multicolumn{3}{c}{\textit{Target designations}} \\
			TIC      & 259168516            & 1 \\
			2MASS    & J19291521+6558279    & 2 \\
			UCAC 4   & 780-032486           & 3 \\
			\gaia\ EDR3 & 2242756094328104576 & 4 \\
			\midrule
			\multicolumn{3}{c}{\textit{Photometry}} \\
			$TESS$	        & 11.040 $\pm$ 0.007   & 1 \\
			$BP$            & 16.27  $\pm$ 0.02    & 4 \\
			$Gaia$	        & 14.61  $\pm$ 0.002   & 4 \\
			$RP$            & 13.38  $\pm$ 0.02    & 4 \\
			$J$	            & 11.637 $\pm$ 0.020   & 2 \\
			$H$	            & 11.137 $\pm$ 0.023   & 2 \\
			$K$	            & 10.821 $\pm$ 0.020   & 2 \\
			WISE 3.4 $\mu$m	& 0.627  $\pm$ 0.023   & 5 \\
			WISE 4.6 $\mu$m	& 10.449 $\pm$ 0.021   & 5 \\
			WISE 12  $\mu$m	& 10.275 $\pm$ 0.047   & 5 \\
			WISE 22  $\mu$m	& 8.23   $\pm$ 0.446   & 5 \\
			\midrule
			\multicolumn{3}{c}{\textit{Astrometry}} \\
			RA  (J2000)     & 19 29 15.21          & 4 \\
			DEC (J2000)     & +65 58 27.72         & 4 \\
			RA PM (mas/yr)  & 56.490 $\pm$ 0.022   & 4 \\
			DEC PM (mas/yr) & -131.659 $\pm$ 0.020 & 4 \\
			Parallax (mas)  & 26.8860 $\pm$ 0.0158 & 4 \\
			\bottomrule
		\end{tabular*}
	\end{center}
 1. \citet{Stassun_2018}, 2. \citet{Cutri_2003}, 3. \citet{Zacharias_2012}, 4. \citet{GaiaCollaboration_2020}, 5. \citet{Cutri_2013}.
		\label{tableTOI-1680}
	
\end{table}
\endgroup

\begingroup
\begin{table}
\caption{Stellar parameters. Parameters in bold are the adopted stellar values in the analyses.}
\begin{center}
\renewcommand{\arraystretch}{1.15}
\begin{tabular*}{\linewidth}{@{\extracolsep{\fill}}l c c@{}}
\toprule
Parameter                   & Value                      & Source                     \\
\midrule
Sp. Type                    & M4.5$\pm$0.5               &  Shane/Kast$^a$            \\
                            & M3.5$\pm$0.5               &  IRTF/SpeX$^b$             \\
\midrule
$T_{\rm eff}/{\rm K}$       & $3191_{-79}^{+81}$         &  SED                       \\
		                      & \textbf{3224$\pm$100}      &  \cite{Mann_2015}          \\
		                    & 3210$\pm$100               &  Stefan-Boltzmann          \\
\midrule
$\mathrm{[Fe/H]}$           &   $+0.04\pm0.20$           &  Shane/Kast$^c$            \\
                            &  \textbf{$-0.32\pm0.13$}   &  IRTF/SpeX $^d$            \\
\midrule
$M_\star/\msun$             & \textbf{0.1800$\pm$0.0044} &  \cite{Mann_2019}          \\
                            & 0.208$\pm$0.010            &  \cite{Mann_2015}          \\ 
                            & 0.1801$\pm$0.021            &  SED                       \\ 
\midrule
$R_\star/\rsun$             & \textbf{0.2130$\pm$0.0064} &  \cite{Mann_2015}          \\
		                      & 0.209$\pm$0.011            &  SED                       \\
\midrule
$L_\star.10^{-3}/\lsun$     & $4.07_{-0.32}^{+0.25}$     &  SED                       \\
                            & $4.353 \pm 0.340$          & $BC_{K}$ \citep{Mann_2015} \\            
\midrule
$\log g_\star / {\rm dex}$  & $5.063_{-0.048}^{+0.047}$  &  SED                       \\
\midrule
$\rho_\star$ / g\,cm$^{-3}$ & 27.40$\pm$2.82             &  $M_\star,R_\star$         \\
                            & $28.4_{-3.8}^{+4.5}$       &  SED                       \\
                            & $27.3_{-2.4}^{+2.3}$       &  Global fit                \\
\bottomrule
\end{tabular*}
\end{center}

$^a$ Classification based on spectral templates from \citet{2007AJ....133..531B} and spectral type/index relations from \citet{1995AJ....110.1838R}, \citet{1999AJ....118.2466M}, \cite{2003AJ....125.1598L} and \citet{2007MNRAS.381.1067R}. \\
$^b$ Classification based on comparison to spectral standards \citep{Cushing2005, Rayner2009} and spectral type/index relations from \citet{Kirkpatrick2010}. \\
$^c$ Metallicity based on measurement of the $\zeta$ index \citet{2007ApJ...669.1235L,2013AJ....145...52M} and calibrations from \citet{2013AJ....145...52M}. \\
$^d$ Metallicity based on measurement of $K$-band Na\,\textsc{i} and Ca\,\textsc{i} doublets and the H2O--K2 index \citep{Rojas-Ayala2012}, and the calibrations of \citet{Mann2014}.
 \label{table:star}
\end{table}
\endgroup

\subsection{Statistical validation}

To statistically validate \tar\,b, we used \texttt{TRICERATOPS}\footnote{\url{https://github.com/stevengiacalone/triceratops}} \citep{Giacalone_2021}, which validates planets by simulating astrophysical false positives arising from gravitationally bound stellar companions, chance-aligned foreground or background stars, and known nearby stars that are blended with the target in the TESS data. It uses a Bayesian tool that incorporates prior knowledge of the target star, planet occurrence rates, and stellar multiplicity to calculate the false positive probability (FPP) and nearby false positive probability (NFPP). The FPP quantity represents the probability that the observed transit is due to something other than a transiting planet around the target star and the NFPP quantity represents the probability that the observed transit originates from a resolved nearby star rather than the target star. \citet{Giacalone_2021} state that for a planet to be statistically validated it must have $\rm FPP < 0.015$ and $\rm NFPP < 0.001$.

We first applied \texttt{TRICERATOPS} to the \TESS\ 2-min-cadence light curve supplied with the contrast curve obtained by the NIRC2 speckle imaging in Sect.~\ref{NIRC2}. The resulting FPP and NFPP values are $0.0018 \pm 0.0001$ and $0.0017 \pm 0.0001$ respectively. The FPP is good enough but NFPP is above the threshold to classify a validated planet \citep{Giacalone_2021}. Only three nearby stars were bright enough and close enough to the target star to cause nearby false positives: TIC 1884271108 ($\Delta T = 6.4$, ${\rm sep} = 6\arcsec$), TIC 259168518 ($\Delta T = 1.3$, ${\rm sep} = 30\arcsec$), and TIC 259168513 ($\Delta T = 2.6$, ${\rm sep} = 37\arcsec$). However, because the event observed by TESS was confirmed to be on-target by our ground-based observations, we were able to rule out these stars as sources of false positives and set NFPP = 0 from the outset. The FPP is then reduced to $0.0001 \pm 0.0001$, which is low enough for validating the planet. Independently, we also used the light curves obtained by the Artemis 1-m and LCO-Hal 2-m telescopes as they present tighter photometric constraints than the TESS data. This was supplied with the same contrast curve mentioned above and without removing any nearby star. This yields FPP and NFPP values generally lower than 0.01 and 0.001 respectively. Therefore, we consider this \TESS\ candidate to be a validated planet.

\subsection{Stellar activity}

\begin{figure}
	\includegraphics[width=\columnwidth]{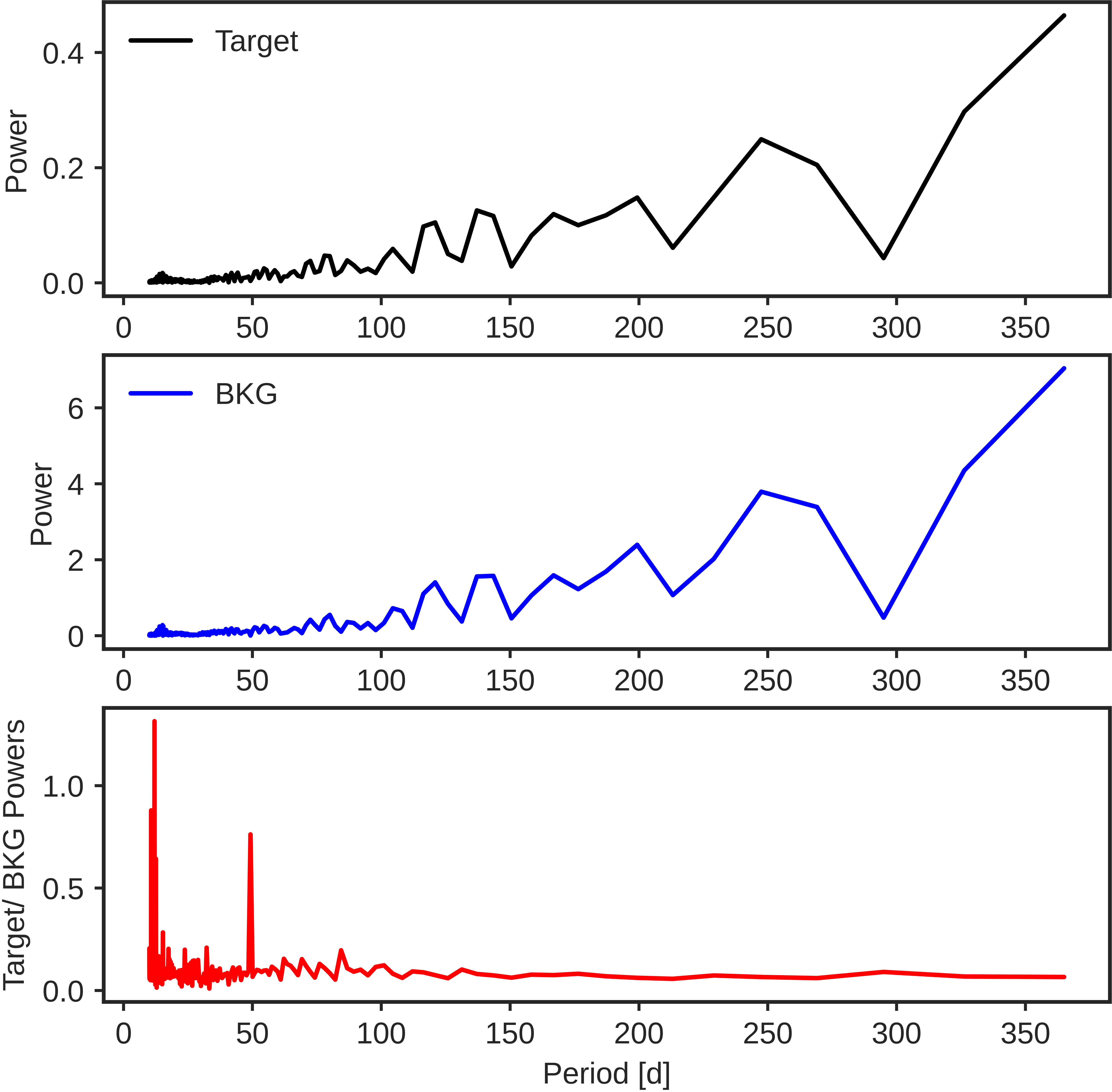}
	\caption{Systematics-insensitive periodogram (SIP) for \tar\,b.
The periodogram is calculated for both the corrected light curve (black line in the top panel)
and the background (BKG) pixels (blue line in the middle panel). The star and the BKG show the same powers. The bottom panel shows the target to the BKG ratio. We do not see any clear peak for the target.}
	\label{fig:SIP}
\end{figure}

With an ecliptic latitude of $\beta = +81.05$ deg, \tar\ is located near the northern ecliptic pole in the \TESS\ CVZ. Targets in the CVZ are highly valuable for extracting long-period rotation rates. We first visually inspected the \TESS\ PDC-SAP light curves and found no hints of rotational modulation nor evidence of flares. We then used the Systematics-Insensitive Periodogram (\texttt{TESS-SIP} \footnote{\url{https://github.com/christinahedges/TESS-SIP}}) to build a periodogram for photometric data from all the 19 sectors. This tool creates a Lomb-Scargle periodogram, while simultaneously detrending \TESS\ systematics using a similar method to that described in \cite{Angus2016} for detrending systematics in the NASA Kepler/K2 dataset. Since the rotational period of the star might be removed by the PDC pipeline, \texttt{TESS-SIP} uses the \TESS\ target pixel file (TPF) data, and apertures assigned by the \TESS\ Pipeline to reproduce a simple aperture photometry (SAP) light curves of the target. We applied this operation on all the 19 observing sectors. Searching for periods between 10 and 365 days, we applied \texttt{TESS-SIP} on the target and on all the background pixels outside of the \TESS\ pipeline aperture in the \TESS\ TPFs for \tar. The SIP powers, presented in Fig.~\ref{fig:SIP}, show a marked similarity between the periodograms of the target and of the background. For comparison, the lower panel shows the ratio of the target to the background powers. We do not see any significant peaks where the target would display power that would be substantially greater than that of the the background. 

We also used \texttt{TESS-SIP} on each sector alone. We did not find any clear stellar rotation signal nor consistency between the periodograms of any sectors. In short, there is no hint of rotational variability detected for \tar\ in the \TESS\ dataset, which is consistent with the lack of detectable H$\alpha$ emission in its optical spectrum.  

Although the TESS CVZ light curves would encompass typical rotation periods for mid-M-dwarfs \citep[0.1 to 140 days; e.g.,][]{2016ApJ...821...93N},  we searched for completeness other ground-based photometric archives. Our target \tar\ is not part of the MEarth sample \citep{2012AJ....144..145B}. We analyzed the  ASAS-SN light curves \citep{2014ApJ...788...48S,2017PASP..129j4502K} for our object that had observations in $V$ and $g$, spanning from June 2012 to June 2023. We do not find rotational modulation. We note that given the faintness of our target in bluer bands (e.g., Gaia$_{BP}$ = 16.27 mag), only $\sim$65\% of the ASAS-SN observations were above the observational limit of each individual observation (with a median limiting magnitude of 16.832 mag in the full light curve). Furthermore, flares cannot be robustly detected in the ASAS-SN data because of its cadence \citep[about two to three days;][]{2018MNRAS.477.3145J}.

\subsection{Transit modeling}

We jointly analyzed the light curves from \TESS\ and ground-based instrumentation described in Sect.~\ref{obs} using the \texttt{EXOFASTv2} \citep{Eastman_2019} software package. We included the \TESS\ photometric data described in Sect.~\ref{TESS:phot} from sectors 14 to 50. We detrended the ground-based light curves for the airmass, as well as for either the background or the half width at half maximum (HWHM) of the PSF. The choice of these parameters was based on the likelihood maximization. Some light curves were detrended only for airmass, especially the partial ones. Table \ref{tab:obs} shows the detrending parameters of each light curve. The detrending was done simultaneously with the transit fitting to ensure a good propagation of the uncertainties on the derived parameters. We fixed the eccentricity to zero assuming the orbit to be circular (see justification below). We set the \texttt{NOMIST} flag to disable the MIST stellar track that constrains the star and, instead, we imposed Gaussian priors on the stellar mass ($0.1800\pm0.0044 \msun$), radius ($0.2100\pm0.0064 \rsun$) and temperature ($3224\pm100 K$) from our determinations reported in Table \ref{table:star} as well as uniform priors on the period ($P = 4.8026 \pm 0.1P$\,d) and transit epoch ($T_{c}\pm P/3$) from the values reported in $ExoFOP$. We set the \texttt{NOCLARET} flag to disable the Claret tables \citep{Claret_2017,Claret&Bloemen_2011} that are used to fit the quadratic limb-darkening parameters  $u_{1}$ and $u_{2}$ and we applied our own gaussian priors computed using the \texttt{PyLDTk} code \citep{LDTK} for each passband (see Table \ref{tab:ld}). TESS light curves are corrected for contamination, but we still fit for dilution of the transit signal in the \TESS\ band due to the neighboring stars using 0$\pm$10\% of the contamination ratio reported in the TIC 259168516 on ExoFOP as Gaussian prior as recommended by \cite{Eastman_2019} to account for any uncertainty in the correction. We ran the EXOFASTv2 analysis until convergence when the Gelman-Rubin statistic (GR) and the number of independent chain draws (Tz) were less than 1.01 and greater than 1000, respectively.

\begin{figure*}
\centering
\includegraphics[scale=0.67]{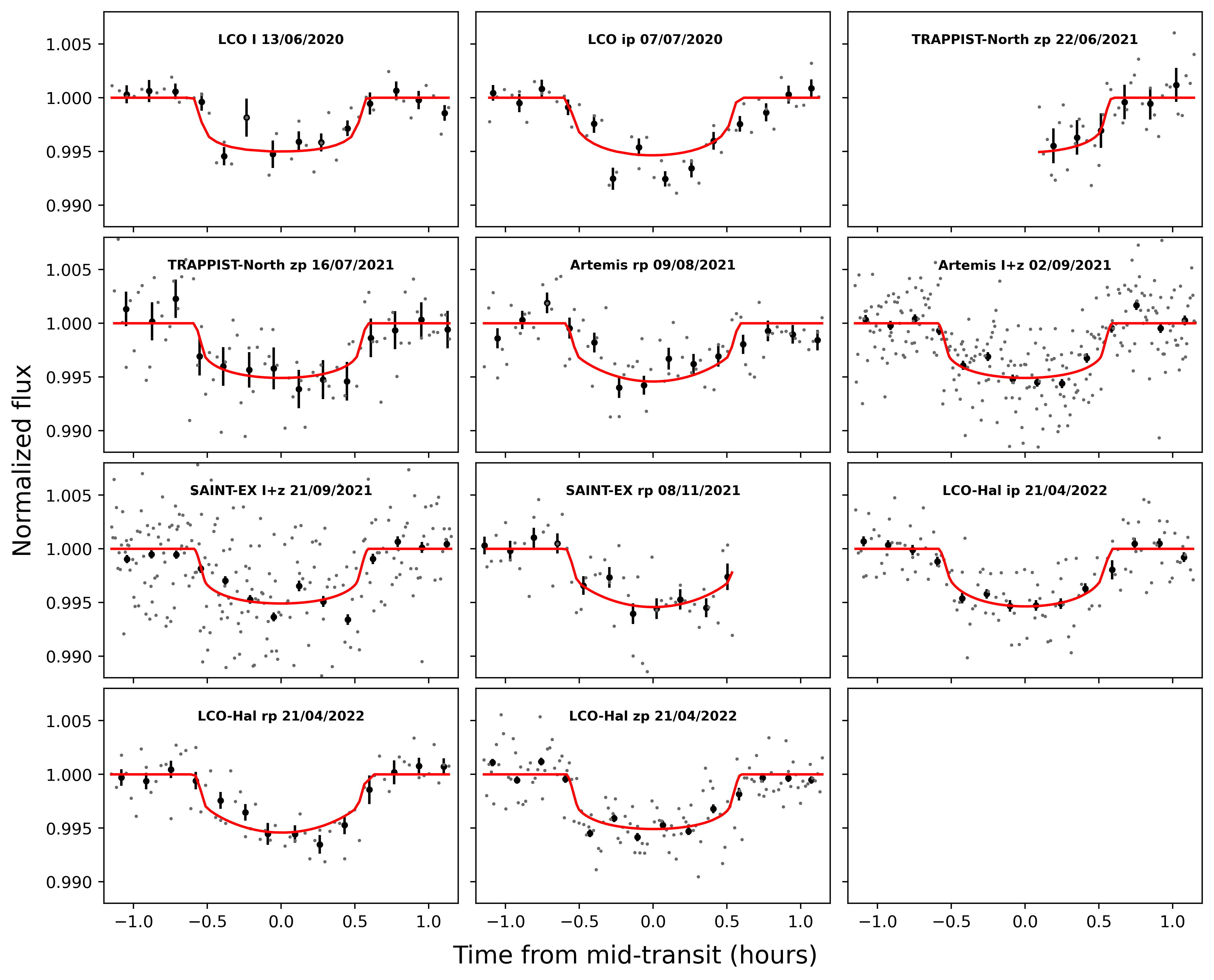}
\caption{Phase-folded ground-based transit light curves of \tar\,b. The unbinned light curves are shown with gray points. The binned points (bin size=10 min) are shown in black with the corresponding error bars. The red line corresponds to the final \texttt{EXOFASTv2} transit model.}
\label{fits}
\end{figure*}

\begin{figure}
	\centering
\includegraphics[width=\columnwidth]{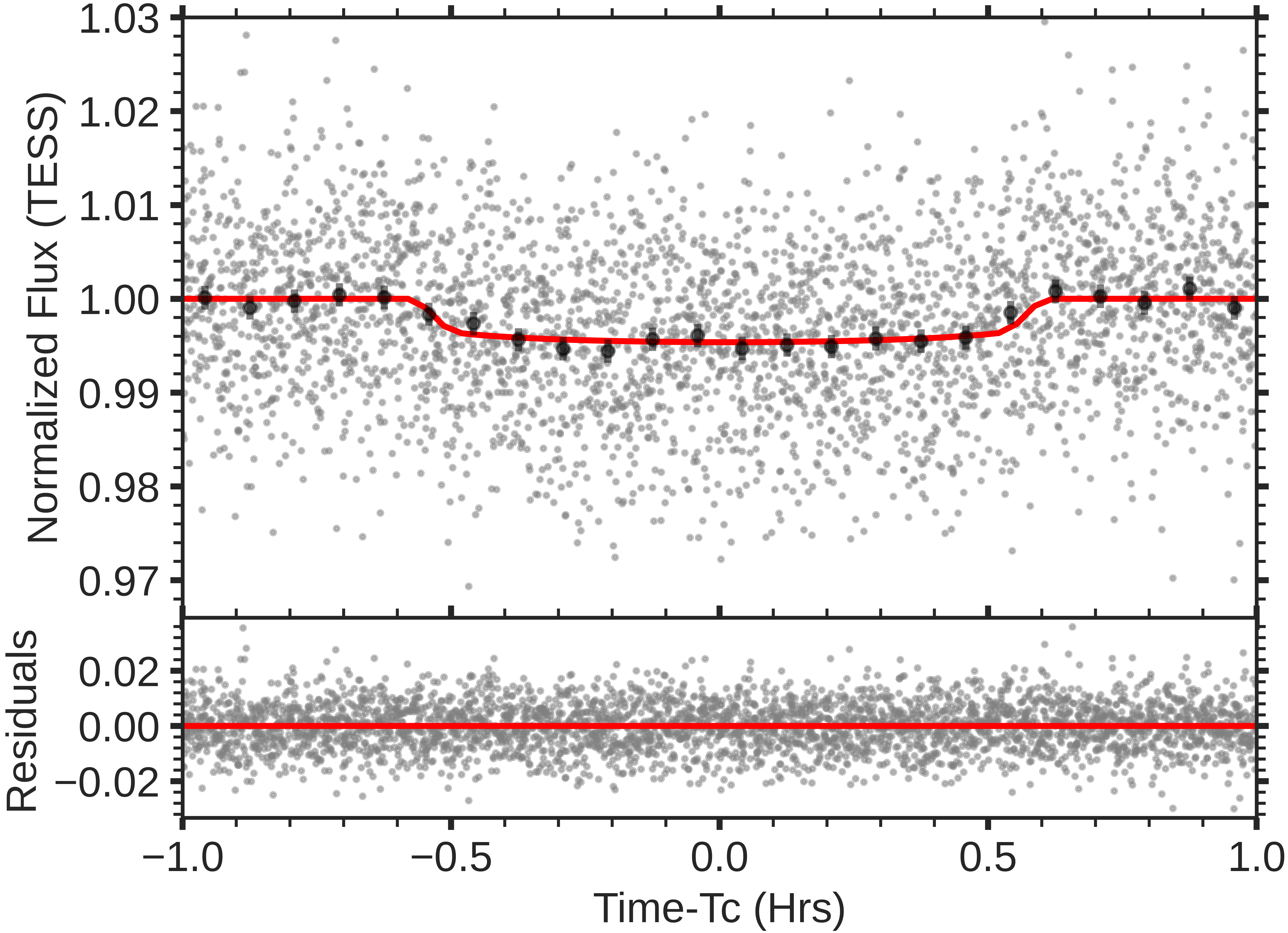}
\caption{\TESS\ light curve folded in phase with the time in hours from mid-transit of \tar\,b. The red solid line represents the best-fit transit model from the final joint-fit. The gray dots are the 2-min \TESS\ data and the black dots are the binned data every phase interval of 5 minutes. The residuals are plotted in the bottom panel.}
\label{fig:TESS_Phased}
\end{figure}

To test the impact of the detrending on our derived parameters, we performed independent analyses of the light curves with another code. We used the \texttt{PyTransit} package \citep{Parviainen_2015MNRAS} and linear models of detrending vectors (FWHM, airmass and background), as done in \cite{Wells2021} and \cite{Schanche2022}. We found good agreement between the fits and therefore continued with \texttt{EXOFASTv2} for the full analysis. The detrended and modeled light curves are presented in Fig.~\ref{fits} and the \TESS\ phase-folded light curve is presented in Fig.~\ref{fig:TESS_Phased}. The results are reported in Table \ref{fit:results}. We also fitted for an eccentric orbit to assess the evidence for orbital eccentricity using photometric-only data. A comparision of the loglikelihood of the two fits appears to favor a circular orbit.

We were concerned that the background level was overestimated and overcorrected in the SPOC pipeline in the northern year 2 sectors (14-26). Fitting only data from sectors 14 to 50 would lead to an overestimation of the planet radius by roughly 2\%. However, this bias is comparable to the error bars on the planet radius and, the inclusion of the dilution term and additional ground-based data significantly mitigates the problem. 

\begin{table}
\caption{Quadratic limb-darkening coefficients used in the photometric joint analysis for each passband.\label{tab:ld}}
\begin{center}
\begin{tabular}{l c c l}
\toprule
Filter &  $u_1$ & $u_2$ \\
\midrule
\TESS\ & $0.2759\pm0.0300$ & $0.3347\pm0.0500$ \\
$z'$   & $ 0.2367\pm0.0277$ & $0.3540\pm0.0529$ \\
$r'$   & $ 0.5211\pm0.0533$ & $0.3247\pm0.0790$ \\
$g'$   & $ 0.5262\pm0.0457$ & $0.3611\pm0.0646$ \\
$Ic$   & $ 0.2940\pm0.0280$ & $0.3729\pm0.0516$ \\
$i'$   & $ 0.3121\pm0.0301$ & $0.3743\pm0.0550$ \\
\hline 
\end{tabular}
\end{center}
\end{table}

{\renewcommand{\arraystretch}{1.4}
\begin{table*}[!htbp] 
\caption{Median values and 68\% confidence intervals for the parameters of \tar\,b obtained using EXOFASTv2.} 
   \begin{center}
{\small %
	\begin{tabular}{llc}
    	\hline
    	\textbf{Parameter} & \textbf{Description} & \textbf{Best-fit value} \\ 
        \hline
        \hline
\smallskip\\\multicolumn{2}{l}{Stellar Parameters:} & \smallskip                            \\

~~~~$M_*$\dotfill                 & Mass (\msun)\dotfill              & $0.1798\pm0.0044$\\
~~~~$R_*$\dotfill                 & Radius (\rsun)\dotfill            & $0.2106^{+0.0061}_{-0.0048}$\\
~~~~$L_*$\dotfill                 & Luminosity (\lsun)\dotfill        & $0.00434^{+0.00062}_{-0.00055}$\\
~~~~$\rho_*$\dotfill              & Density (cgs)\dotfill             & $27.2^{+2.0}_{-2.3}$\\
~~~~$\log{g}$\dotfill             & Surface gravity (cgs)\dotfill     & $5.046^{+0.027}_{-0.021}$\\
~~~~$T_{\rm eff}$\dotfill         & Effective Temperature (K)\dotfill & $3225^{+99}_{-100}$\\

~~~~\smallskip\\\multicolumn{2}{l}{Planetary Parameters:}              &   \smallskip         \\
~~~~$P$\dotfill           & Orbital period (days)\dotfill              & $4.8026345^{+0.0000040}_{-0.0000039}$\\
~~~~$R_P$\dotfill         & Planet radius (\re)\dotfill                & $1.466^{+0.063}_{-0.049}$\\
~~~~$T_C$\dotfill         & Transit time     (\bjdtdb)\dotfill         & $2459013.84254^{+0.00037}_{-0.00039}$\\
~~~~$a$\dotfill           & Semi-major axis (AU)\dotfill               & $0.03144^{+0.00025}_{-0.00026}$\\
~~~~$i$\dotfill           & Inclination (Degrees)\dotfill              & $89.58^{+0.26}_{-0.22}$\\
~~~~$T_{\rm eq}$\dotfill        & Equilibrium temperature$^{a}$ (K)\dotfill& $404\pm14$\\
~~~~$R_P/R_\star$\dotfill       & Planet-to-star radius ratio\dotfill            & $0.0638\pm0.0015$ \\
~~~~$a/R_\star$\dotfill         & Semi-major axis in stellar radii\dotfill & $32.04^{+0.74}_{-0.93}$\\
~~~~$\delta$\dotfill            & $\left(R_P/R_\star\right)^2$\dotfill     & $0.00407^{+0.00019}_{-0.00018}$\\
~~~~$\delta_{\rm I}$\dotfill    &Transit depth in $I$ (fraction)\dotfill   & $0.00472\pm 0.00020$\\
~~~~$\delta_{\rm i'}$\dotfill   &Transit depth in $i'$ (fraction)\dotfill  & $0.0066^{+0.0023}_{-0.0015}$\\
~~~~$\delta_{\rm r'}$\dotfill   &Transit depth in $r'$ (fraction)\dotfill  & $0.0066^{+0.0020}_{-0.0014}$\\
~~~~$\delta_{\rm z'}$\dotfill   &Transit depth in $z'$ (fraction)\dotfill  & $0.00503^{+0.00098}_{-0.00065}$\\
~~~~$\delta_{\rm TESS}$\dotfill &Transit depth in \TESS\ (fraction)\dotfill& $0.00507^{+0.0012}_{-0.00070}$\\
~~~~$\tau$\dotfill     & Ingress/egress transit duration (days)\dotfill    & $0.00312^{+0.00026}_{-0.00016}$\\
~~~~$T_{14}$\dotfill   & Total transit duration (days)\dotfill             & $0.04941^{+0.00069}_{-0.00065}$\\
~~~~$T_{FWHM}$\dotfill & FWHM transit duration (days)\dotfill              & $0.04625^{+0.00072}_{-0.00069}$\\
~~~~$b$\dotfill        & Transit impact parameter\dotfill                  & $0.23^{+0.11}_{-0.14}$\\
~~~~$\fave$\dotfill    & Incident Flux (\fluxcgs)\dotfill     &$0.00600^{+0.00084}_{-0.00075}$\\
~~~~$d/R_*$\dotfill    & Separation at mid-transit\dotfill &$32.04^{+0.74}_{-0.93}$\\
\smallskip\\\multicolumn{2}{l}{Predicted Parameters:}& \smallskip\\
~~~~$M_P$\dotfill      & Planet mass$^{b}$ (\me)\dotfill          &$3.18^{+1.1}_{-0.69}$\\
~~~~$K$\dotfill        & RV semi-amplitude (m/s)\dotfill   &$3.78^{+1.3}_{-0.82}$\\
~~~~TSM\dotfill        & Transmission spectroscopy metric\dotfill  &7.82\\
~~~~$M_P/M_*$\dotfill  & Planet-to-star mass ratio\dotfill          &$0.000053^{+0.000019}_{-0.000011}$\\

~~~~$\rho_P$\dotfill   & Planet mean density (cgs)\dotfill    &$5.5^{+1.9}_{-1.2}$\\
\hline
\end{tabular}
}%
\end{center}
$^{a}$The equilibrium temperature corresponds to a case with null albedo and an 100\% efficient heat recirculation. $^{b}$The planetary mass is estimated from the planetary radius from transit using the \cite{Chen_Kipping2017} mass--radius relation. The radial velocity semi-amplitude ($K$) is predicted using the estimated mass.\label{fit:results}
\end{table*}}

\section{Discussion} \label{discus}

\subsection{\tar\,b and the radius valley}

Studies performed by \cite{Fulton_2017} on the Kepler small exoplanets have identified a radius valley roughly from 1.5 to 2 \re, separating rocky super-Earths and gaseous sub-Neptunes around Sun-like stars. A low-radius peak at $1.3\,\re$ corresponds to high-density super-Earths and a high-radius peak at $2.4 \re$ corresponds to low-density sub-Neptunes with significant primordial H/He atmospheres. This gap is considered as possible transition region between rocky and icy "super-Earth" and "mini-Neptunes." \citet{Cloutier_Menour_2020} showed that the radius valley persists for low-mass stars (i.e., $M \lesssim 0.65$ \rsun). 

Two contrasting theories have been presented to explain the radius valley. The first is gas-poor formation model which proposes that the radius valley is a feature intrinsic to the exoplanet population from formation onward  \citep{Luque_2021,Lee_2014, Lee_2016,Lee_2021}. Specifically, some planets are formed with extended H/He envelopes, whereas the population of rocky planets is formed later in a gas-poor environment after the gas is dissipated from the protoplanetary disk. The second is thermally driven atmospheric mass loss \citep{Lopez_2013, Owen_2013, Jin_2014, Chen_2016}, which proposes that the radius gap is formed through evolution after the gas accretion phase. That is, planets are formed with gaseous envelopes and some of them experience atmospheric escape through two scenarios: 1) photoevaporation \citep{Lopez_2013, Owen_2013, Jin_2014, Chen_2016} triggered by energetic EUV and X-ray flux from the host star in the first $\sim$100 Myrs \citep{Owen_2013} of the system and 2) core-powered mass-loss \citep{Ginzburg_2016} triggered by the energy emergent from the cooling planetary core in a Gyr timescale \citep{Ginzburg_2018}. 

On the contrary, a recent study performed by \cite{Luque&palle2022} further suggests that there is a density gap, but not a radius gap, separating rocky planets and water-rich worlds with no planets with intermediate composition. Using a sample of 34 exoplanets with well-characterized densities around M-dwarf stars, they identified three populations: rocky planets ($\rho=0.94\pm0.13~\rho_{\oplus}$), water-rich planets ($\rho=0.47\pm0.05~\rho_{\oplus}$), and gas-rich planets ($\rho=0.24\pm0.04~\rho_{\oplus}$). These study findings favor the pebble accretion model \citep{2020Venturini,2020Brugger} as the main mechanism for forming small planets around M dwarfs, where rocky planets are formed within the ice line while water-rich planets are formed beyond the ice line and then migrated inwards. However, as for previous studies, the sample of exoplanets used in this study is not large enough to draw firm conclusions. 

\begin{figure}
	\centering
	\includegraphics[width=\columnwidth]{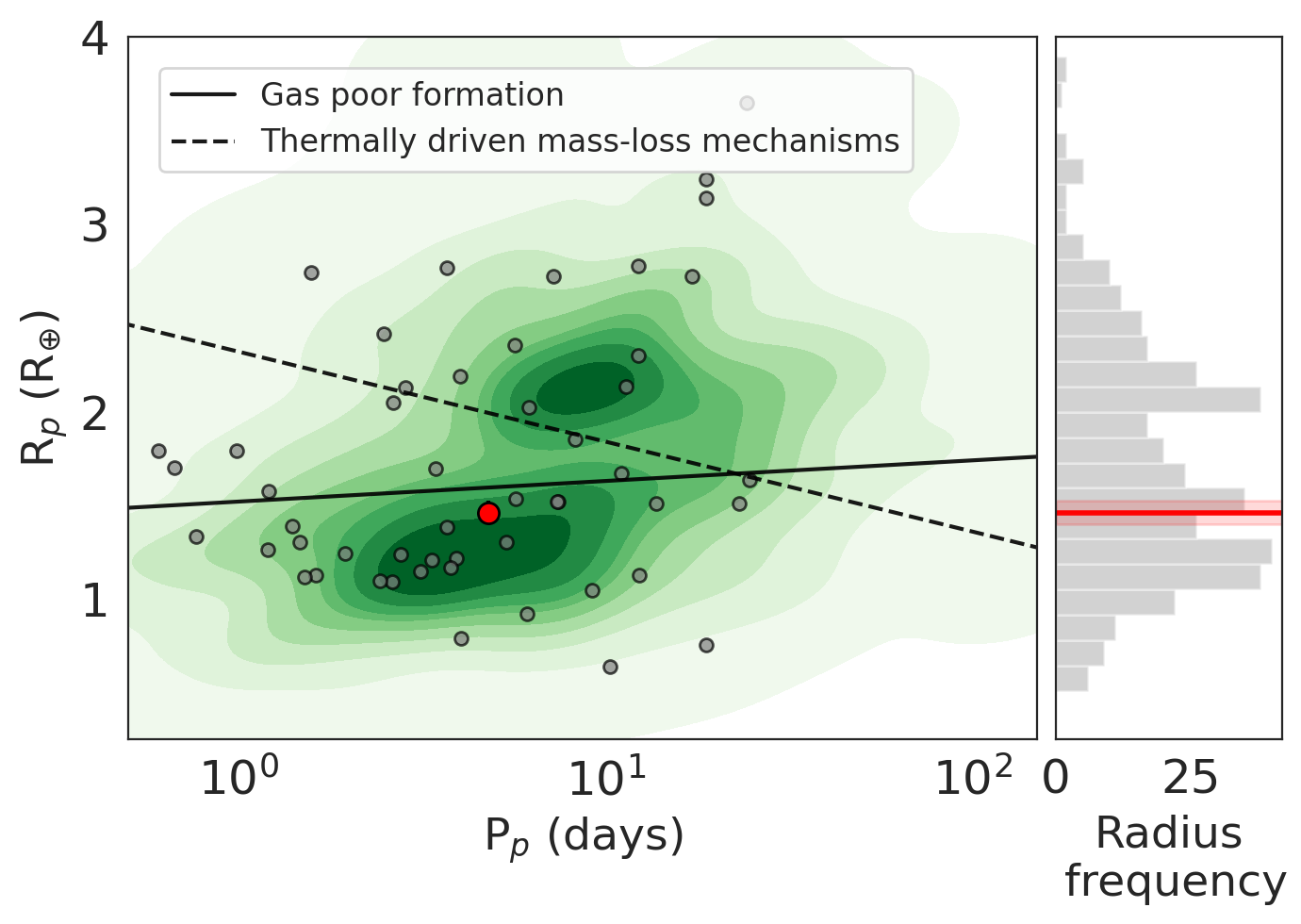}
	\caption{Distribution of planet radii and orbital periods for all confirmed small planets hosted by low-mass stars (M$_{*} \lesssim 0.65 \msun$). The solid and dashed lines represent the predicted locations of the radius valley based on the gas-poor formation model and thermally driven photoevaporation models, respectively. The red dot represents the planet \tar\,b. The 1D radius distribution, with the location of the planet \tar\,b, is shown on the right panel.}
	\label{TOI-1680_RadValley}
\end{figure}

Figure.~\ref{TOI-1680_RadValley} shows the current period--radius diagram of all known exoplanets with precise radius measurements orbiting M dwarfs. The empirical locations of the radius valley for FGK stars as predicted by thermally driven photoevaporation (dashed line) given by \cite{VanEylen_2018} and for low-mass stars as predicted by gas-poor formation (solid line) given by \cite{Cloutier_Menour_2020} are also displayed. With a radius of $1.466^{+0.063}_{-0.049} \re$ and orbital period of $4.8026343\pm0.0000030$ days, \tar\,b is located close to the center of the super-Earth population where the two models predict the location of small rocky planets. With a future planetary mass determination, \tar\,b can join the growing sample of small planets with precise bulk densities around M dwarfs.

\subsection{Prospects for a radial velocity follow-up} \label{RV_}

The precise mass determination of \tar\,b would allow us to better constrain the detectability of a possible atmosphere and better locate the planet in the radius-density gap. High-precision radial velocity (RV) measurements will not only allow us to constrain the planetary mass, but also constrain its orbit such as its eccentricity, which may shed some light onto the dynamical history of the system. 
\tar\,b has a radius of $1.466^{+0.063}_{-0.049} \re$. Thus, we expect a RV semi-amplitude of $3.78^{+1.3}_{-0.82}\,\mathrm{m\,s^{-1}}$, assuming a circular orbit and a mass of $3.18^{+1.1}_{-0.69}\,\me$, as predicted from the mass--radius relation of \cite{Chen_Kipping2017}.

Not many high-precision spectrographs in the northern hemisphere are capable of detecting such a shallow signal from a faint target (V = 15.87 mag and J = 11.63 mag). Many typical planet finders, mounted on 2--4\,m class telescopes such as CARMENES \citep{Quirrenbach_2020S}, HARPS-N \citep{Cosentino_2012}, NEID \citep{Schwab_2016} or EXPRES \citep{Jurgenson_2016SPIE} have limiting magnitudes that are brighter than V=15.87. MAROON-X at the 8.1\,m Gemini North telescope \citep{2020SPIE11447E..1FS} has shown to reach the necessary precision for faint M-dwarf host stars \citep{2022SPIE12184E..1GS}. Despite the high declination, it would be possible to reach a S/N of about 40 in the red arm after a 15\,min exposure. Assuming a stellar activity level of $\rm < 1.5\,m\,s^{-1}$, this would allow for an overall precision of $0.7\,\mathrm{m\,s^{-1}}$ precision with about 70 spectra. Thus, it will be possible to measure the planet mass with state-of-the-art instrumentation at $\rm 5\,\sigma$ precision by investing about 29\,h of telescope time on a 8-m-class telescope.

\subsection{Potential for atmospheric characterization}

We assess the potential for atmospheric characterization of \tar\,b with \JWST\ using the transmission spectroscopy metric \citep[TSM,][]{Kempton_2018}. The TSM quantifies the expected S/N in the transmission spectrum of a given planet with a cloud-free atmosphere. Analytically, it is expressed as:
$$
TSM = S\times\frac{R_{p}^{3}T_{eq}}{M_{p}R_{*}^{2}}\times 10^{\frac{-m_{j}}{5}},
$$\\
where R$_{p}$ and M$_{p}$ are the planetary radius and mass in Earth units, $R_{*}$ is the stellar radius in Solar radii, $T_{eq}$ is the equilibrium temperature of the planet in K and $m_{J}$ is the apparent magnitude of the star in the J band. Also, $S$ is a scale factor whose value depends on the planetary radius range.

\begin{figure}
    \centering
    \includegraphics[width=\columnwidth]
    {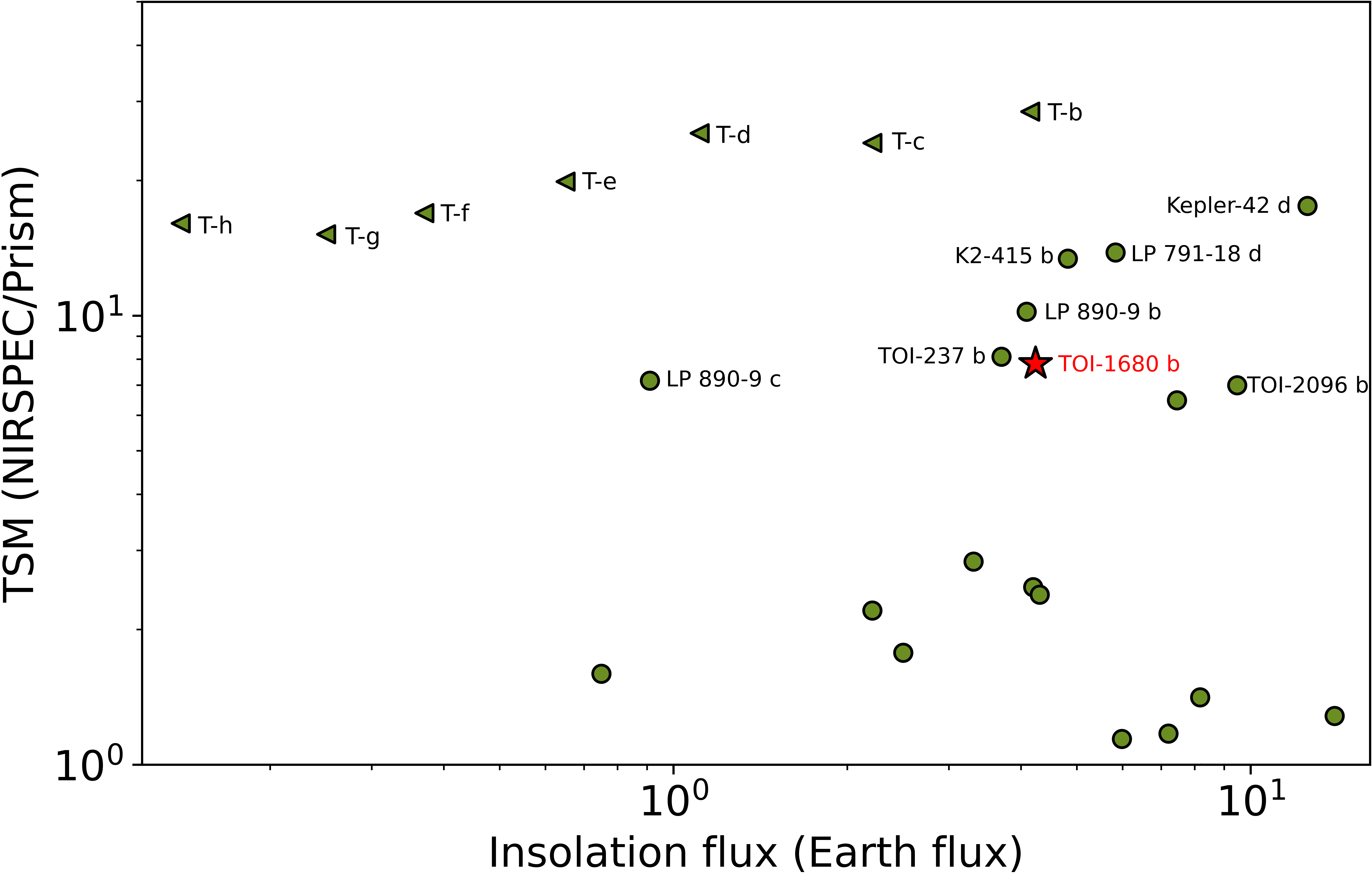}
	\caption{Transmission Spectroscopy Metric for known planets that can be observed with the \textit{JWST} NIRSPEC/Prism. The sample is limited to exoplanets with equilibrium temperatures below 500\,K and radii smaller than 1.5\,\re\ and orbiting stars fainter than $J{=}10.5$. The x-axis is limited to 16 times Earth insolation flux. \tar\,b is labeled and highlighted with a star. Data were taken from NASA Exoplanet Archive on June 25, 2023.}
	\label{TSM}
\end{figure}

\tar\,b is a cool (T$_{eq} <$ 500 K) super-Earth (R < 1.5\re). With J=11.6, the host star is within reach of the \textit{JWST} NIRSpec/PRISM (0.6--5.3 $\mu$m) instrument \citep{Jakobsen2022}, which cannot observe stars brighter than J=10.5 without saturation. We used the TSM to assess the suitability of all cool (T$_{eq} <$ 500 K) planets with radii smaller than $1.5 \re$ orbiting stars fainter than J=10.5 for atmospheric studies with this instrument. This sample of exoplanets contains 63 targets. We used the empirical mass--radius relation of \cite{Chen_Kipping2017} to estimate the mass of planets that do not have mass measurements as is the case for \tar\,b. TSM values of these planets are shown in Fig.~\ref{TSM}. We found that \tar\,b has a TSM=7.82 which indicate that it could be a suitable target for transmission studies with the \JWST\ NIRSpec/PRISM instrument. Specifically, amongst 63 targets, \tar\,b ranks as the thirteenth most amenable target for these studies. It follows all the TRAPPIST-1 planets \citep{Gillon_2017}, Kepler-42 d \citep{Muirhead_2012}, K2-415 b \citep{Hirano2023},  LP 791-18 d \citep{Peterson2023Natur}, LP 890-9 b \citep{Delrez2022} and TOI-237 b\citep{Waalkes_2021AJ}. Moreover, the TSM is based on ten hours of \JWST\ observing time and with an ecliptic latitude of $\beta = +81.05 \deg$. \tar\,b is located near the \JWST\ CVZ, and it has the advantage of being observable for about 250 days per year, which encourages its atmospheric characterization.

\section{Conclusion} \label{concl}

We have reported the discovery and initial characterization of \tar\,b, a super-Earth orbiting a faint mid-M dwarf (V=15.87). We used the combination of 2-min cadence \TESS\ observations from 19 sectors, ground-based photometry, high-angular-resolution imaging and spectroscopic observations to validate its planetary nature. Joint analyses of \TESS\ and ground-based data yielded a planetary radius of $1.466^{+0.063}_{-0.049}$\,\re, an orbital period of $4.8026345^{+0.0000040}_{-0.0000039}$\,days and an equilibrium temperature of $404\pm14$\,K. According to the transmission spectroscopy metric (TSM) of \cite{Kempton_2018}, \tar\,b could be a promising candidate for atmospheric characterization with the \JWST. However, a stronger prediction of the expected S/N waits for direct mass measurement from radial velocity observations, which could be done with the MAROON-X instrument at the 8.1 m Gemini North telescope.

\begin{acknowledgements}

This research received funding from the European Research Council (ERC) under the European Union's Horizon 2020 research and innovation programme (grant agreement n$^\circ$ 803193/BEBOP).
B.V.R. thanks the Heising-Simons Foundation for support.
TRAPPIST is funded by the Belgian Fund for Scientific Research (Fond National de la Recherche Scientifique, FNRS) under the grant FRFC 2.5.594.09.F, with the participation of the Swiss National Science Fundation (SNF). 
The ULiege's contribution to SPECULOOS has received funding from the European Research Council under the European Union's Seventh Framework Programme (FP/2007-2013) (grant Agreement n$^\circ$ 336480/SPECULOOS), from the Balzan Prize and Francqui Foundations, from the Belgian Scientific Research Foundation (F.R.S.-FNRS; grant n$^\circ$ T.0109.20), from the University of Liege, and from the ARC grant for Concerted Research Actions financed by the Wallonia-Brussels Federation. 
This work is based upon observations carried out at the Observatorio Astron\'omico Nacional on the Sierra de San Pedro M\'artir (OAN-SPM), Baja California, M\'exico. SAINT-EX observations and team were supported by the Swiss National Science Foundation (PP00P2-163967 and PP00P2-190080), 
 the Centre for Space and Habitability (CSH) of the University of Bern,  the National Centre for Competence in Research PlanetS, supported by the Swiss National Science Foundation (SNSF),  and UNAM PAPIIT-IG101321.
The postdoctoral fellowship of KB is funded by F.R.S.-FNRS grant T.0109.20 and by the Francqui Foundation.
This work makes use of observations from the LCOGT network. Part of the LCOGT telescope time was granted by NOIRLab through the Mid-Scale Innovations Program (MSIP). MSIP is funded by NSF.
This research has made use of the Exoplanet Follow-up Observation Program (ExoFOP; DOI: 10.26134/ExoFOP5) website, which is operated by the California Institute of Technology, under contract with the National Aeronautics and Space Administration under the Exoplanet Exploration Program.
Funding for the TESS mission is provided by NASA's Science Mission Directorate. KAC acknowledges support from the TESS mission via subaward s3449 from MIT.
Resources supporting this work were provided by the NASA High-End Computing (HEC) Program through the NASA Advanced Supercomputing (NAS) Division at Ames Research Center for the production of the SPOC data products.
This paper is based on observations made with the MuSCAT3 instrument, developed by the Astrobiology Center and under financial supports by JSPS KAKENHI (JP18H05439) and JST PRESTO (JPMJPR1775), at Faulkes Telescope North on Maui, HI, operated by the Las Cumbres Observatory.This work is partly supported by JSPS KAKENHI Grant Number JP18H05439
and JST CREST Grant Number JPMJCR1761.
This work was partially supported by a grant from the Erasmus+ International Credit Mobility program (M.~Ghachoui).

MG is F.R.S-FNRS Research Director. LD is an F.R.S.-FNRS Postdoctoral Researcher.

J.d.W. and MIT gratefully acknowledge financial support from the Heising-Simons Foundation, Dr. and Mrs. Colin Masson and Dr. Peter A. Gilman for Artemis, the first telescope of the SPECULOOS network situated in Tenerife, Spain. F.J.P. acknowledges financial support from the grant CEX2021-001131-S funded by MCIN/AEI/ 10.13039/501100011033.

This publication benefits from the support of the French Community of Belgium in the context of the FRIA Doctoral Grant awarded to MT.

\end{acknowledgements}
\bibliographystyle{aa}
\bibliography{toi1680.bib}

\clearpage

\onecolumn
\begin{appendix}

\section{Target pixel files of \tar} \label{appendix}

\begin{figure*}[hbt!]

\centering

\includegraphics[width=0.245\textwidth]{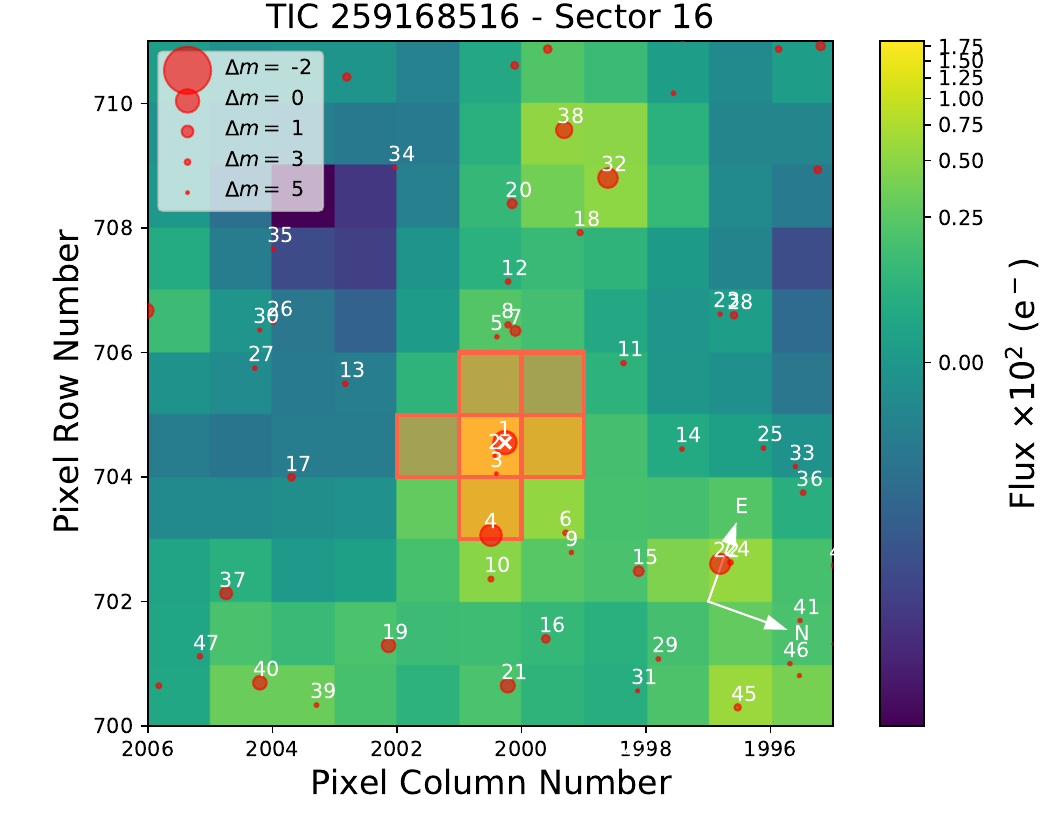}
\includegraphics[width=0.245\textwidth]{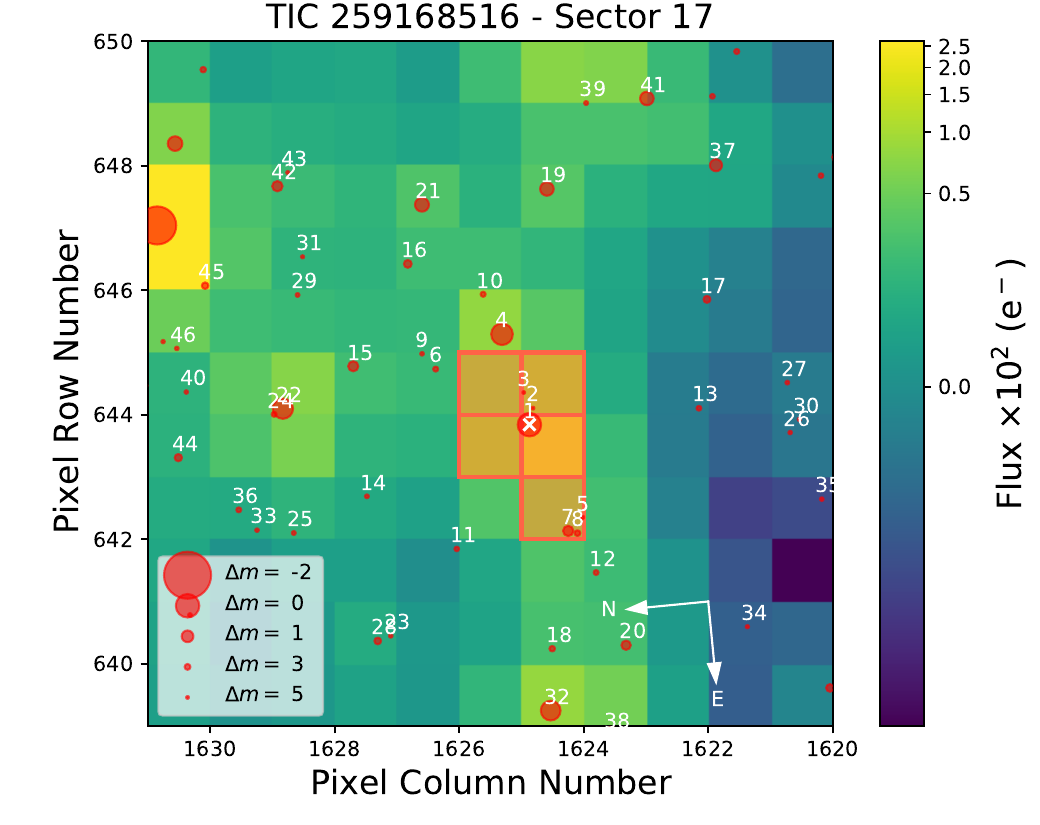}
\includegraphics[width=0.245\textwidth]{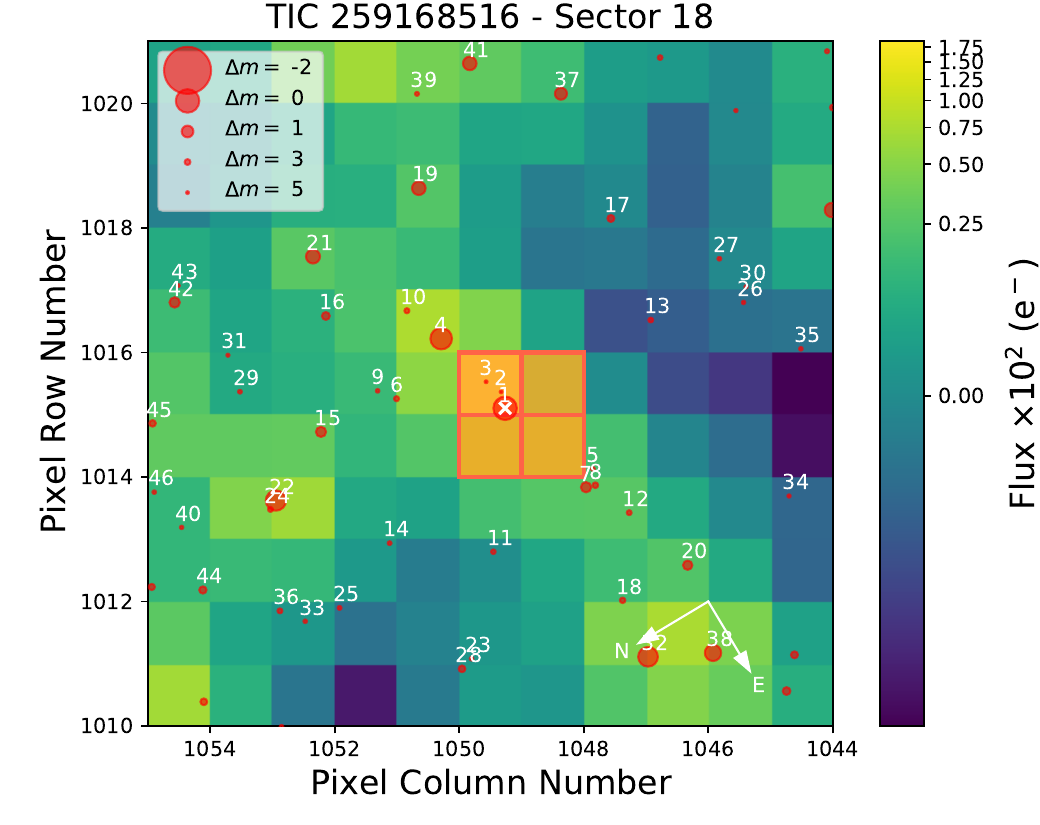}

\includegraphics[width=0.245\textwidth]{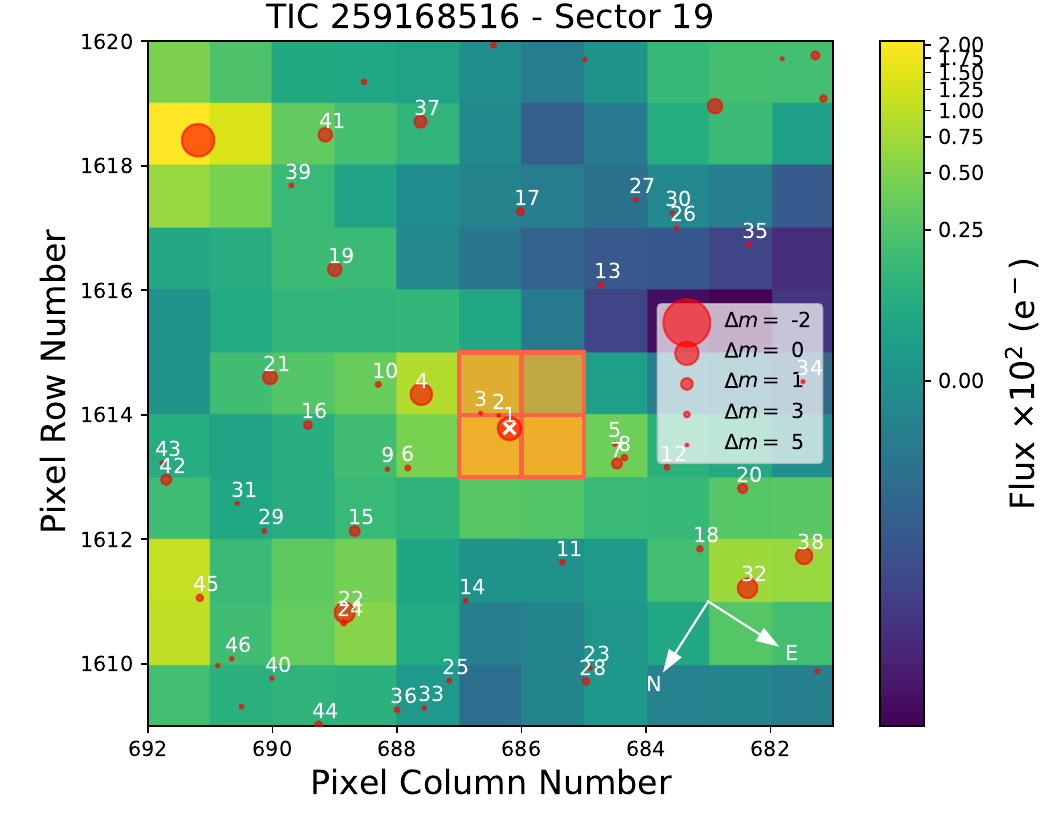}
\includegraphics[width=0.245\textwidth]{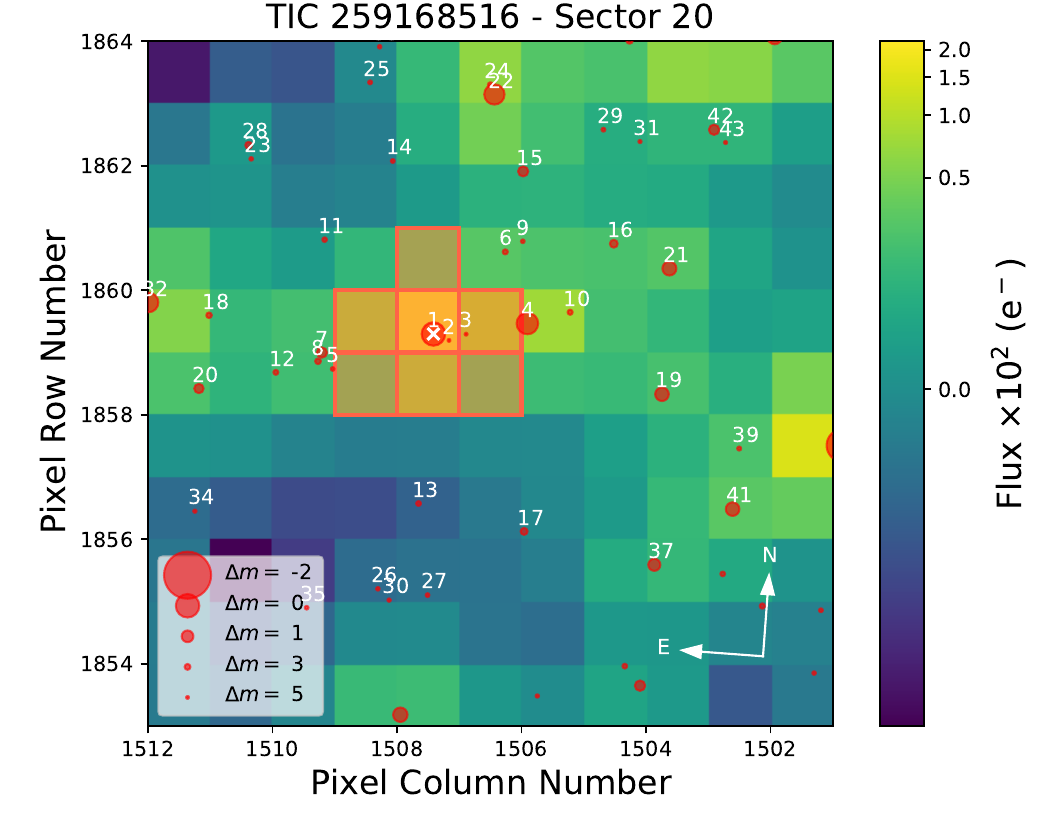}
\includegraphics[width=0.245\textwidth]{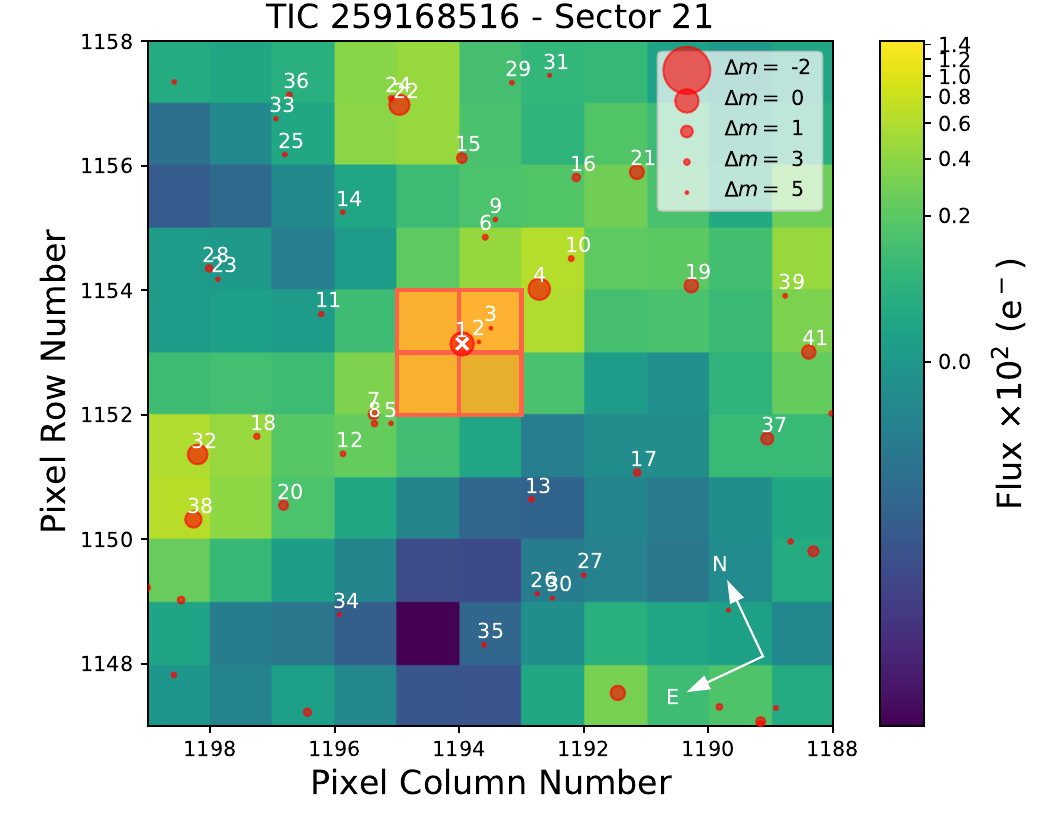}

\includegraphics[width=0.245\textwidth]{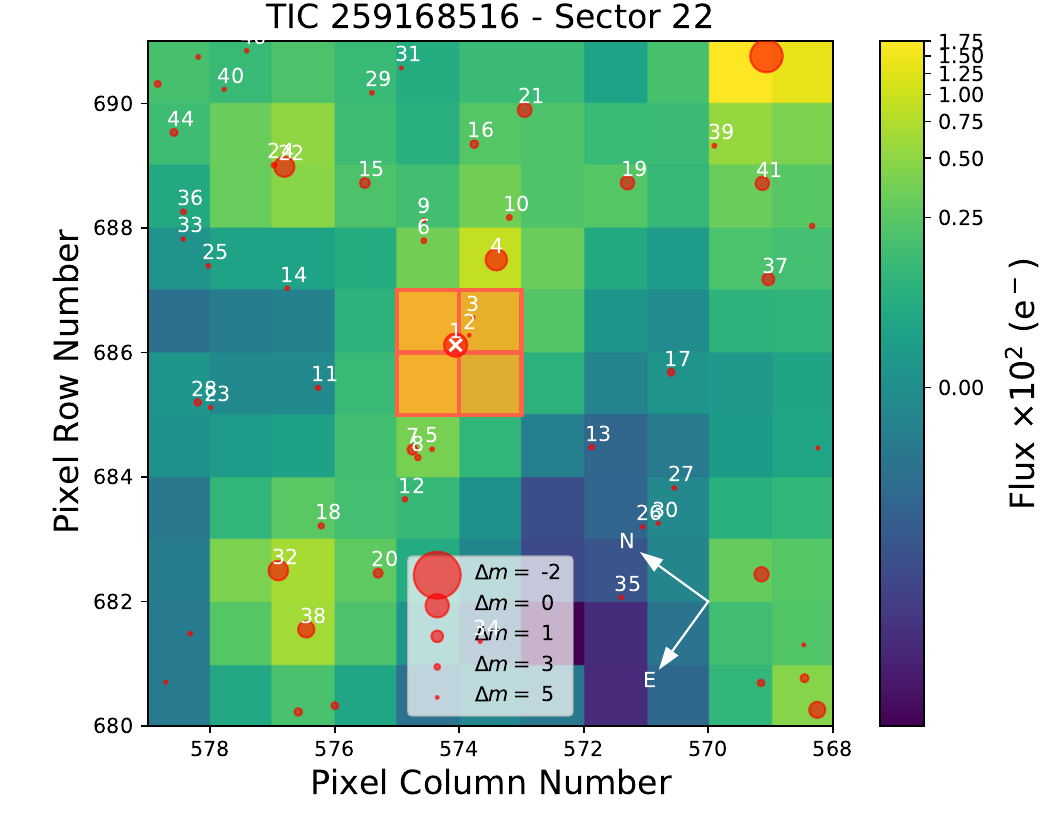}
\includegraphics[width=0.245\textwidth]{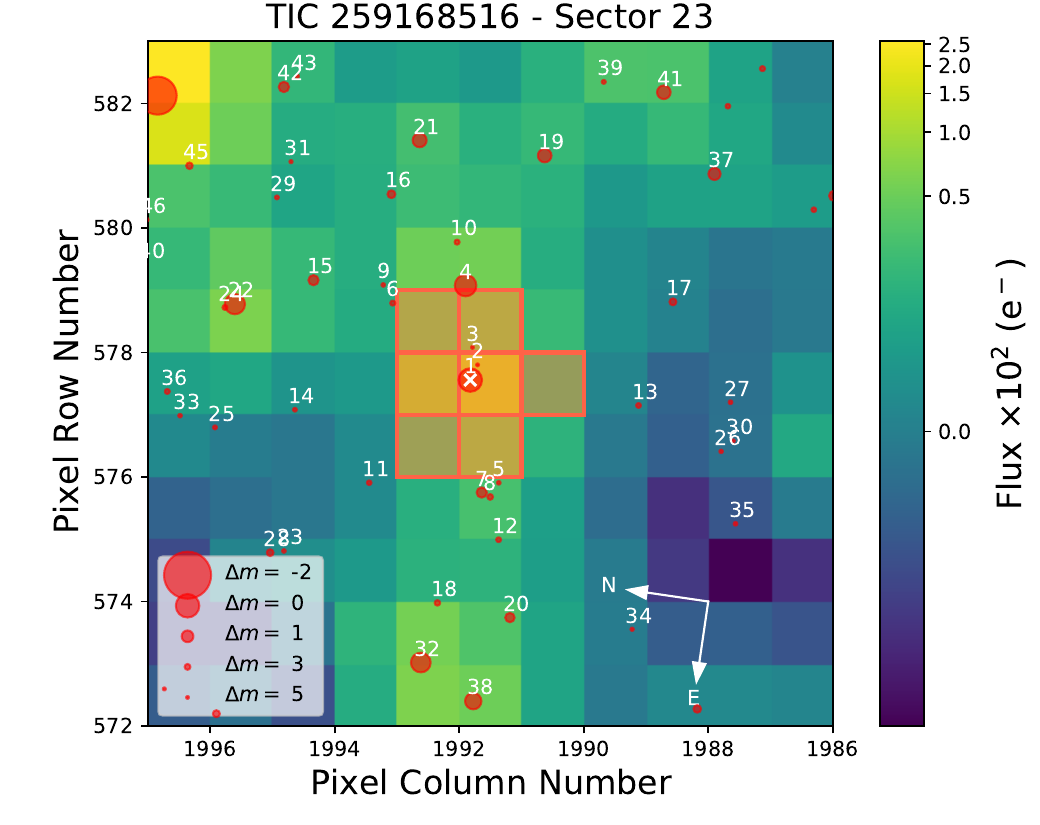}
\includegraphics[width=0.245\textwidth]{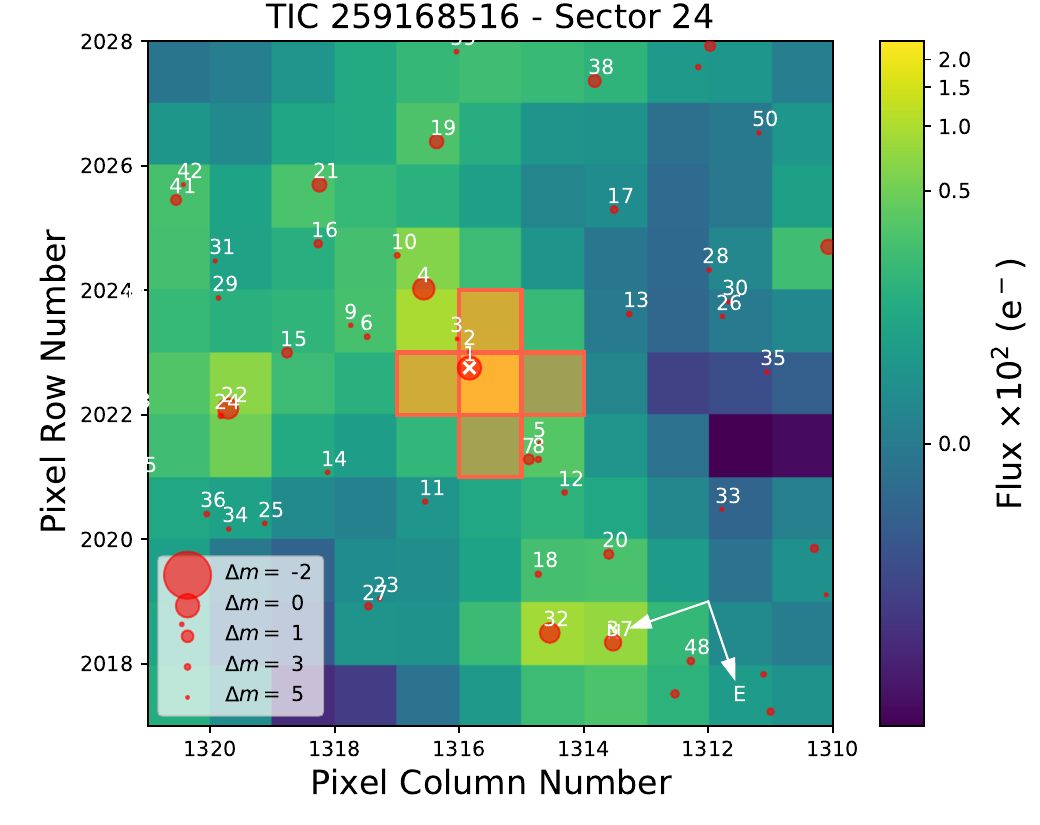}

\includegraphics[width=0.245\textwidth]{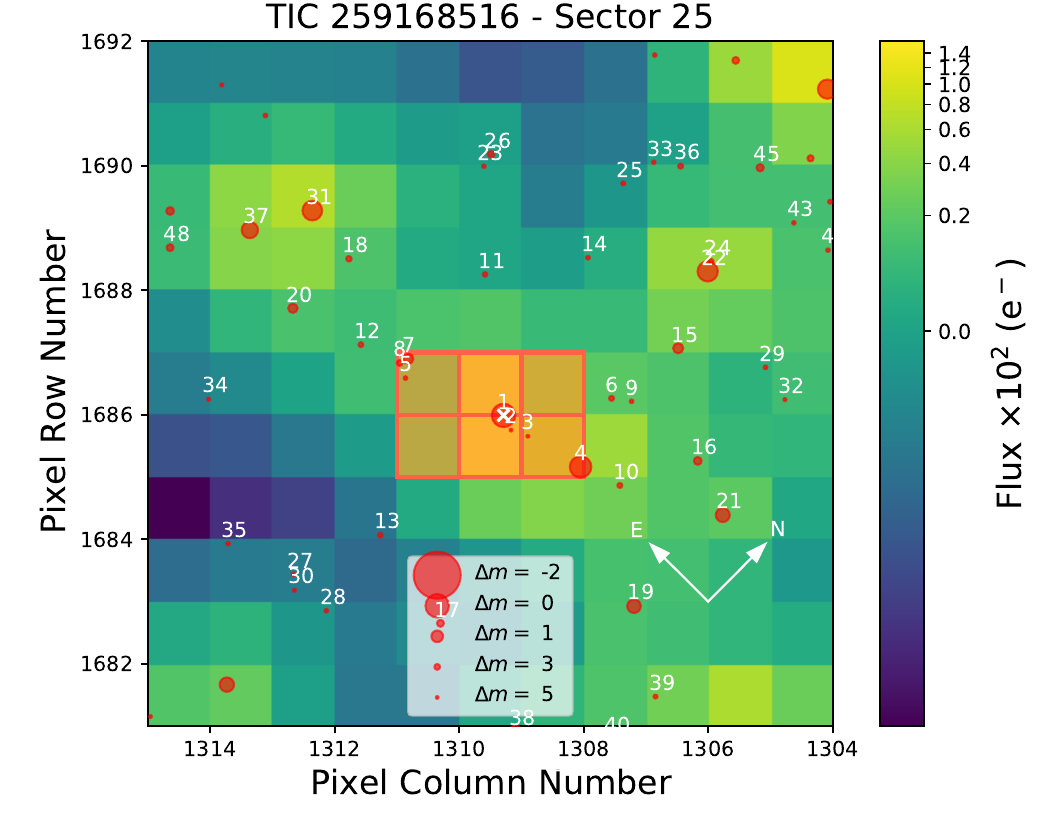}
\includegraphics[width=0.245\textwidth]{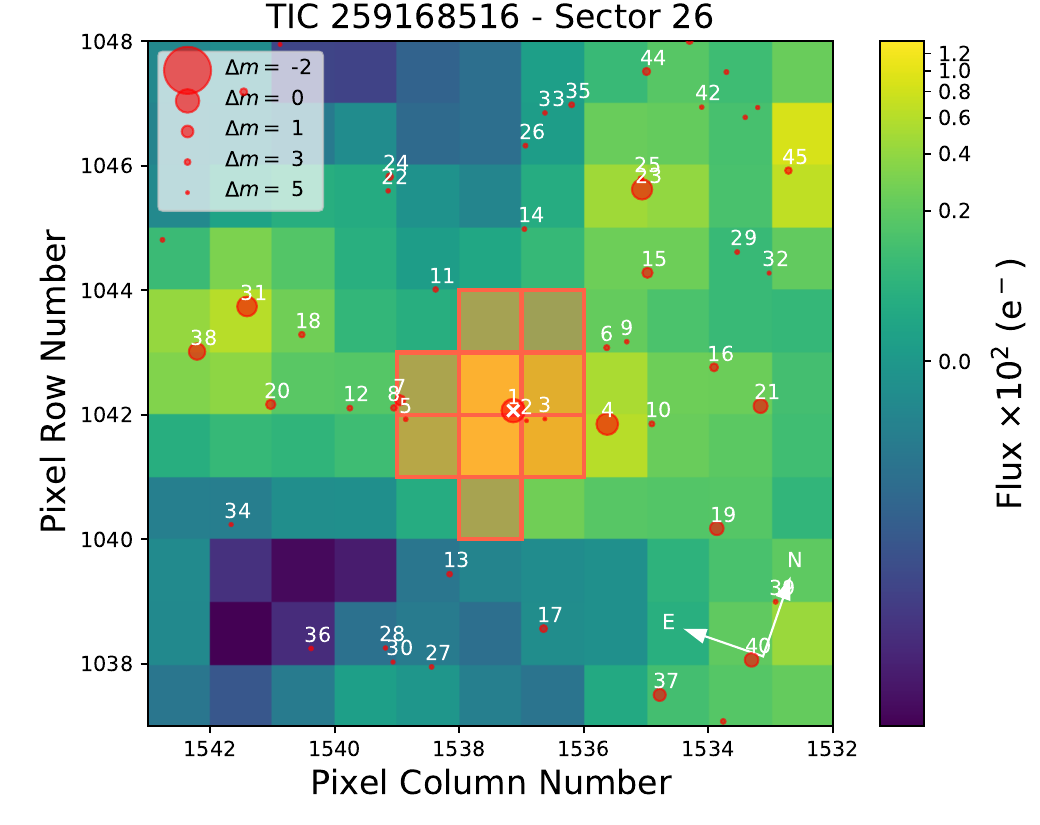}
\includegraphics[width=0.245\textwidth]{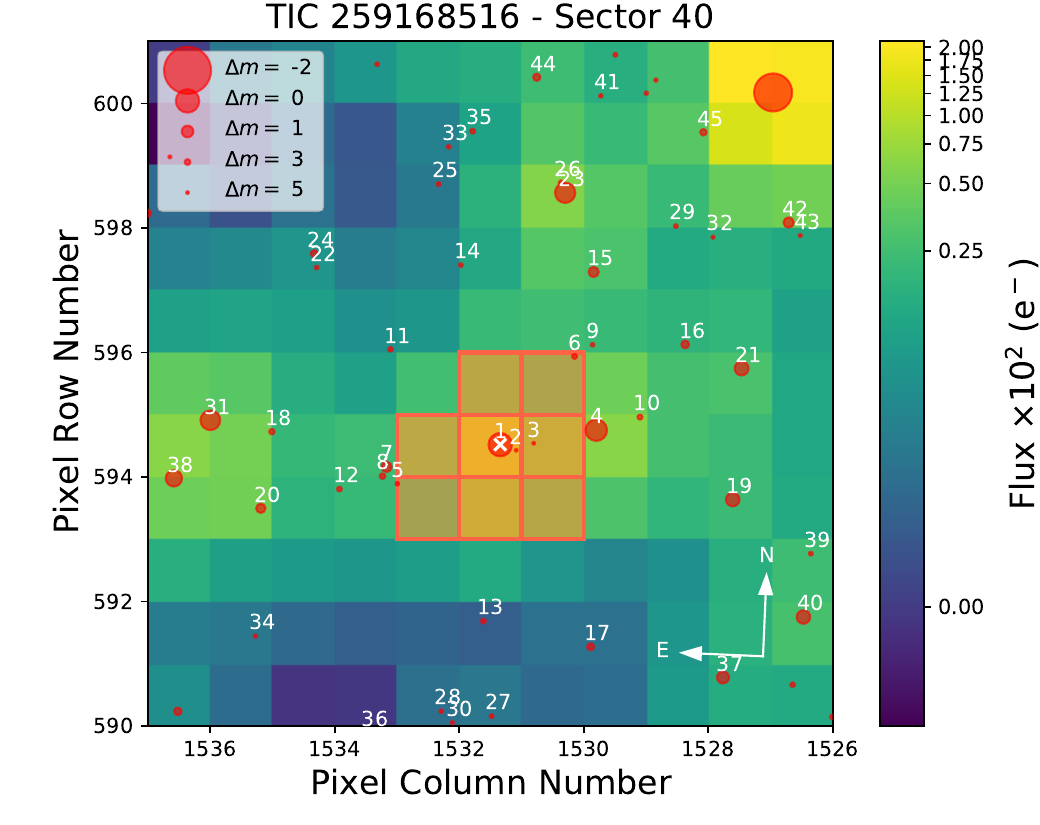}

\includegraphics[width=0.245\textwidth]{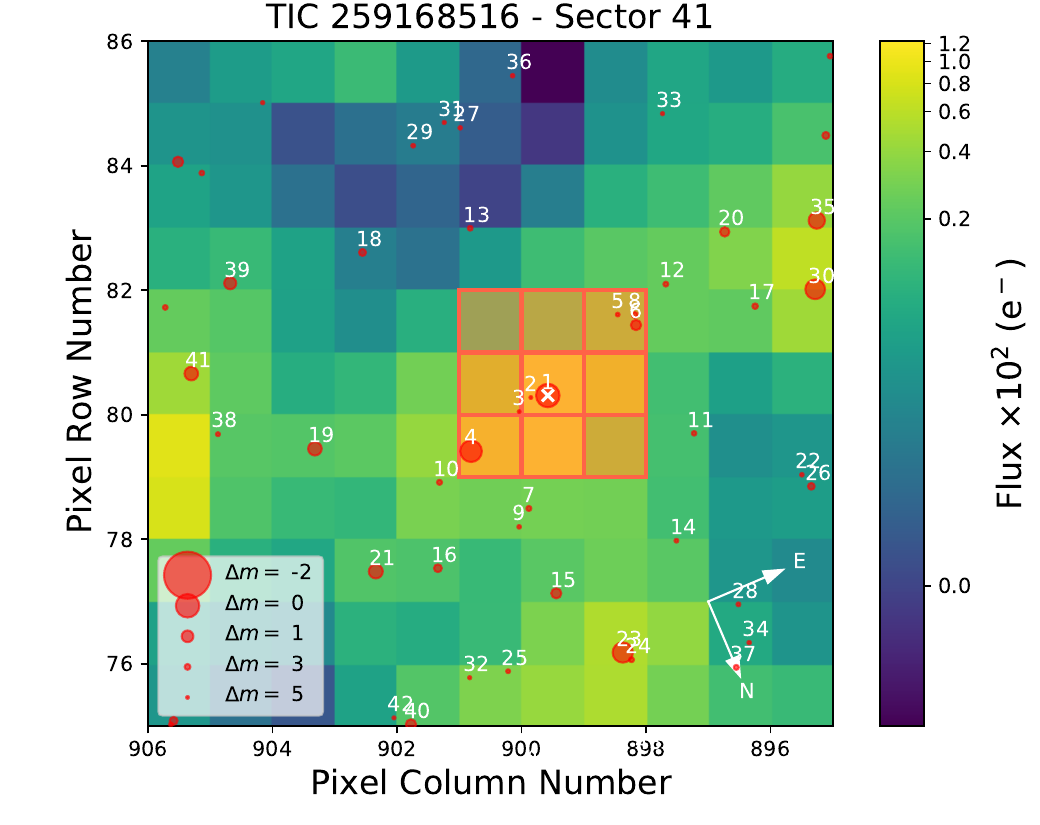}
\includegraphics[width=0.245\textwidth]{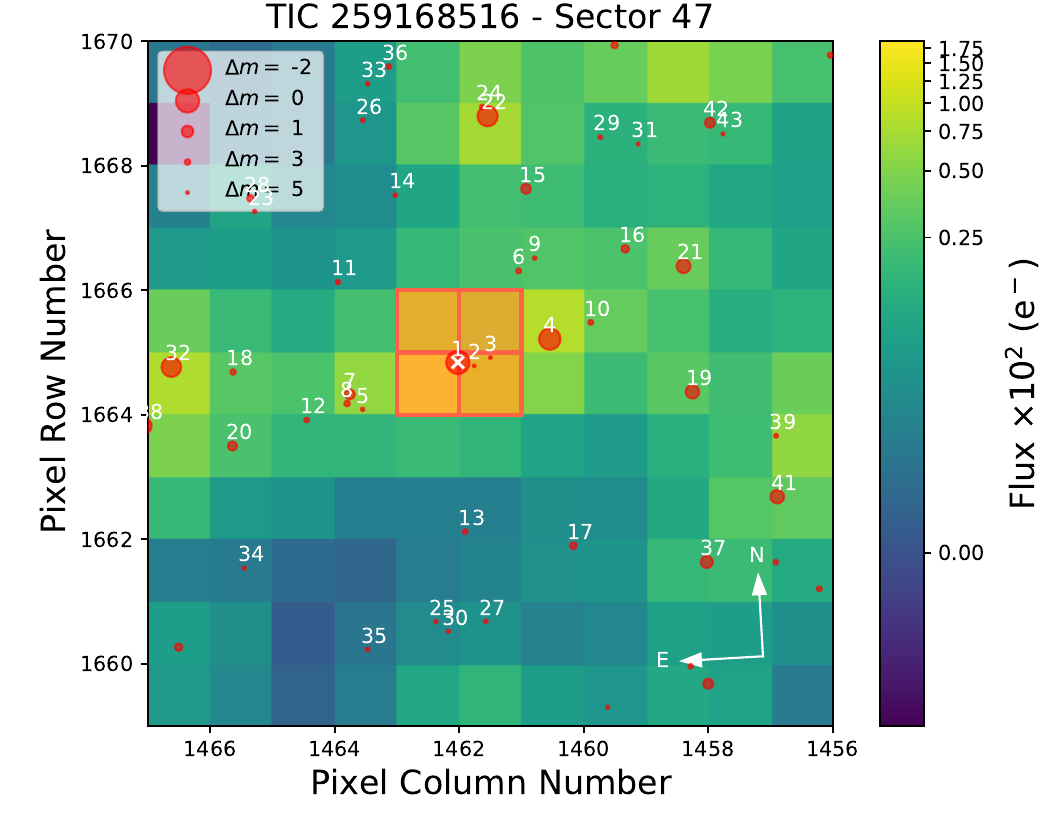}
\includegraphics[width=0.245\textwidth]{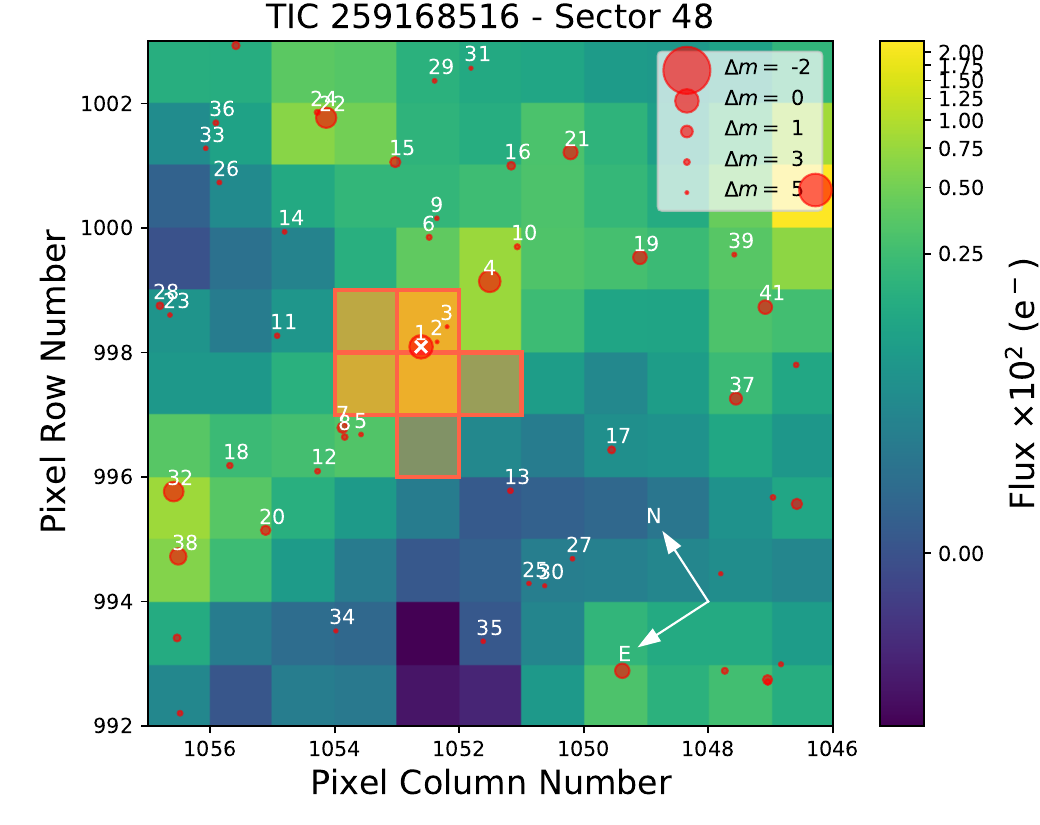}

\includegraphics[width=0.245\textwidth]{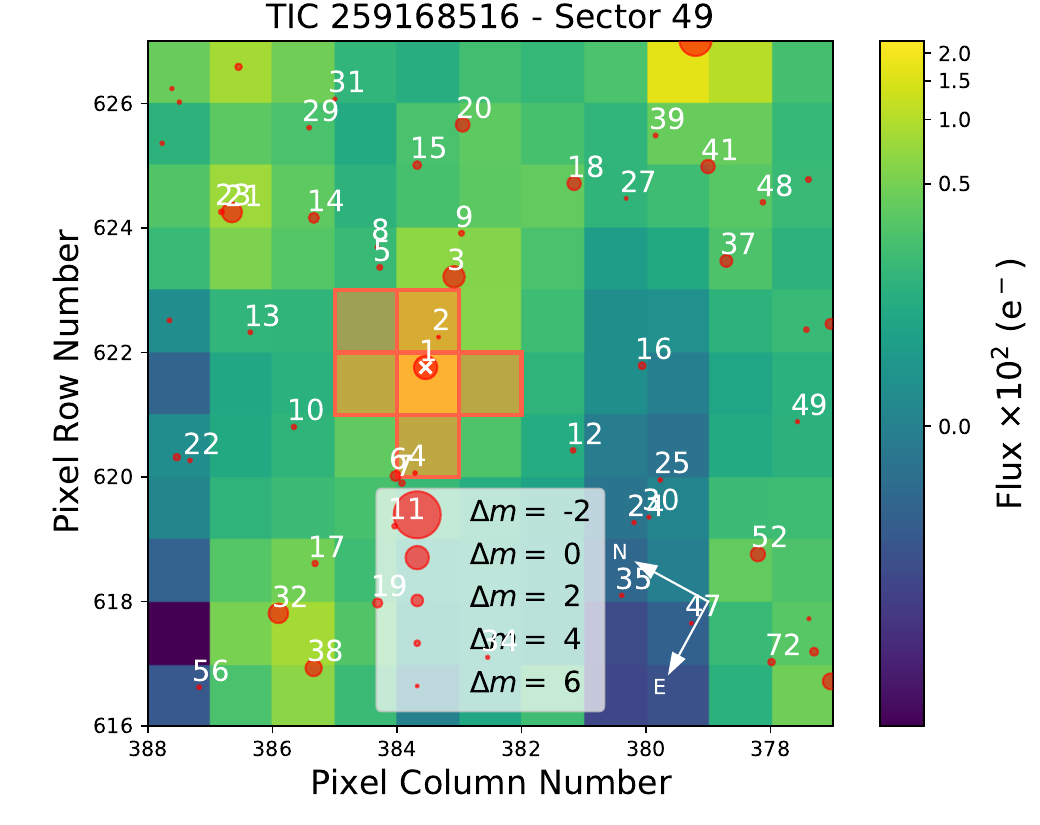}
\includegraphics[width=0.245\textwidth]{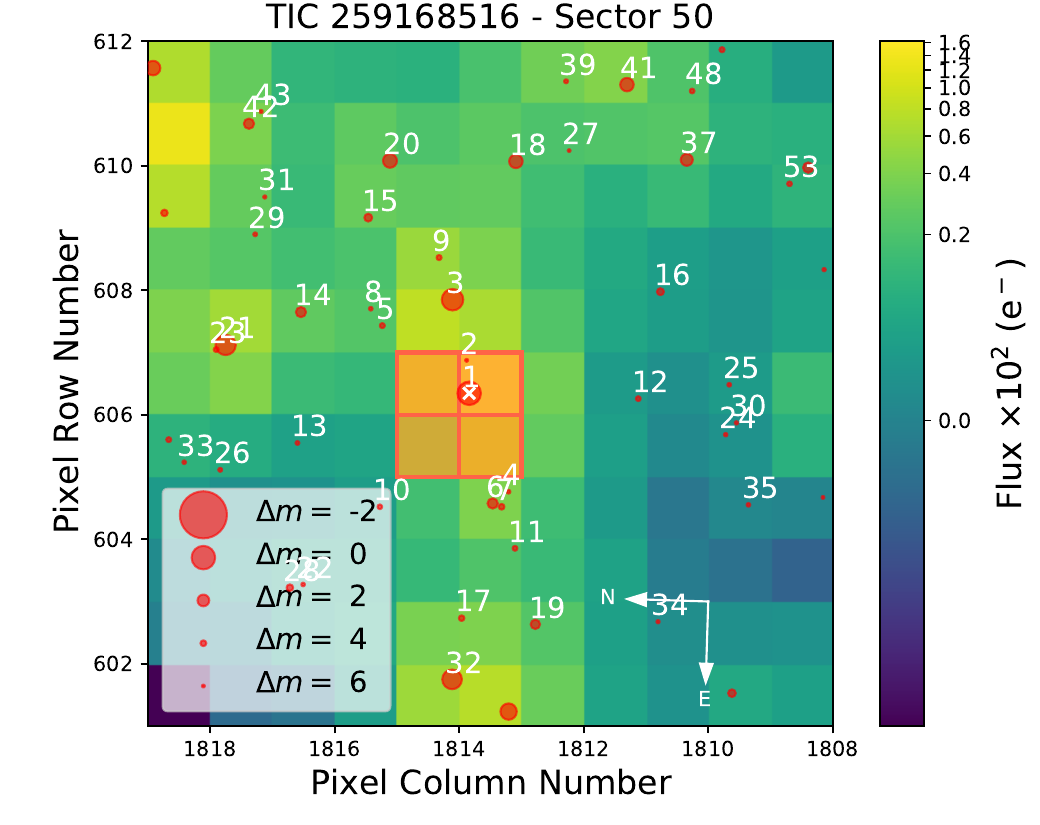}

\caption{\TESS\ target pixel file images of \tar\ observed in Sectors 16, 17, 18, 19, 20, 21, 22, 23, 24, 25, 26, 40, 41, 47, 49, and 50, shown on the left. The red circles show the sources in the field identified by the \gaia\ DR2 catalogue with scaled magnitudes. The position of the targets is indicated by white crosses and the mosaic of orange squares show the mask used by the pipeline to extract the SPOC photometry. These plots were created with \texttt{tpfplotter} \citep{Aller_2020A&A}.} 

\end{figure*}

\end{appendix}

\end{document}